\documentclass[twocolumn]{aastex63}

\usepackage{txfonts} 
\usepackage{natbib}	 
\usepackage{ulem} 
\usepackage{txfonts,epsfig,graphicx,url,twoopt} 
\usepackage{float}
\usepackage{capt-of} 
\newcommand{\RNum}[1]{\uppercase\expandafter{\romannumeral 
#1\relax}}
\usepackage{multirow}
\hypersetup{colorlinks=true,citecolor=blue} 
\bibpunct{(}{)}{;}{a}{}{,}
\newcommand{\adeg}[1]{{#1}$^{\circ}$}
\newcommand\gaia{\textit{Gaia}}
\newcommand\local{\textit{local}}

\received{\today}
\revised{}
\accepted{}

\shorttitle{Evolved Stars in the Galactic Foreground \RNum{1}}
\shortauthors{Quiroga-Nu\~nez et al.}

\graphicspath{{./}{figures/}}

\begin{document}

\title{Characterizing the Evolved Stellar Population in the Galactic Foreground \RNum{1}: \\ Bolometric Magnitudes, Spatial Distribution and P-L Relations}% with~{\gaia}}

\correspondingauthor{Luis Henry Quiroga-Nu\~nez}
\email{lquiroga@nrao.edu}
\author[0000-0002-9390-955X]{Luis Henry Quiroga-Nu\~nez}
\altaffiliation{Jansky Fellow of the National Radio Astronomy Observatory}
\affiliation{National Radio Astronomy Observatory, Array Operations Center, Socorro, NM 87801, USA}
\affiliation{University of New Mexico, Department of Physics and Astronomy, Albuquerque, NM 87131, USA}
\affiliation{Leiden University, Leiden Observatory, Leiden, 2300RA, The Netherlands}
\affiliation{Joint Institute for VLBI ERIC (JIVE), Dwingeloo, 7990AA, The Netherlands}

\author[0000-0002-0230-5946]{Huib Jan van Langevelde}
\affiliation{Joint Institute for VLBI ERIC (JIVE), Dwingeloo, 7990AA, The Netherlands}
\affiliation{Leiden University, Leiden Observatory, Leiden, 2300RA, The Netherlands}

\author[0000-0003-3096-3062]{Lor\'ant O. Sjouwerman}
\affiliation{National Radio Astronomy Observatory, Array Operations Center, Socorro, NM 87801, USA}

\author[0000-0003-0615-1785]{Ylva M. Pihlstr\"om}
\altaffiliation{Adjunct astronomer at the National Radio Astronomy Observatory}
\affiliation{University of New Mexico, Department of Physics and Astronomy, Albuquerque, NM 87131, USA}

\author[0000-0002-7419-9679]{Anthony G. A. Brown}
\affiliation{Leiden University, Leiden Observatory, Leiden, 2300RA, The Netherlands}

\author[0000-0003-0427-8387]{R. Michael Rich}
\affiliation{University of California,
 Department of Physics and Astronomy, Los Angeles, CA 90095, USA}

\author[0000-0002-3019-4577]{Michael C. Stroh}
\affiliation{Northwestern University, Center for Interdisciplinary Exploration and Research in Astrophysics, Evanston, IL 60201, USA}

\author[0000-0002-8069-8060]{Megan O. Lewis}
\affiliation{University of New Mexico, Department of Physics and Astronomy, Albuquerque, NM 87131, USA}
\affiliation{National Radio Astronomy Observatory, Array Operations Center, Socorro, NM 87801, USA}

\author{Harm J. Habing}
\affiliation{Leiden University, Leiden Observatory, Leiden, 2300RA, The Netherlands}

\begin{abstract}
Radio campaigns using maser stellar beacons have provided crucial information to characterize Galactic stellar populations. Currently, the Bulge Asymmetries and Dynamical Evolution (BAaDE) project is surveying infrared (IR) color-selected targets for SiO masers. This provides a sample of evolved stars that can be used to study the inner, optically obscured Galaxy using line of sight velocities and possibly VLBI proper motions. In order to use the BAaDE sample for kinematic studies, the stellar population should be characterized. In this study, the BAaDE targets have been cross-matched with IR (2MASS) and optical ({\gaia}) samples. By exploring the synergies of this cross-match together with~{\gaia} parallaxes and extinction maps, the local ($d < 2$ kpc) AGB stars can be characterized. We have defined a \textit{BAaDE-Gaia} sample of 20,111 sources resulting from cross-matching BAaDE targets with IR and optical surveys. From this sample, a~{\local} sample of 1,812 evolved stars with accurate parallax measurements, confirmed evolved stellar evolution stage, and within 2 kpc distance around the Sun was selected, for which absolute (bolometric) magnitudes are estimated. The evolved stellar population with Gaia counterparts that are variable seems to be predominantly associated with AGB stars with moderate luminosity ($1,500^{+3,000}_{-500} \ L_\odot$) and periods between 250 and 1,250 days.
\end{abstract}


\section{Introduction}

The characterization of Galactic stellar populations is a key ingredient to understand the structural~\citep[see e.g.,][]{Reid2019a}, chemical~\citep[see e.g.,][]{Ibata2017} and dynamical~\citep[see e.g.,][]{Martinez-Medina2017} evolution of the Milky Way, and indeed, its assembly through past merger events~\citep[e.g.,][]{Gomez2012a}. Typically, this is done by combining information on the spatial and kinematic distribution of a stellar population with an assessment of its age and origin~\citep[e.g.,][]{Mackereth2017}. As the~{\gaia} mission~\citep{Brown2018,Lindegren2018} delivers more accurate, reliable data in each data release (DR), it is revolutionizing our understanding of the assembly of the Galaxy. Many recent results demonstrate that mergers have been frequent over the history of the Milky Way~\citep{Antoja2018,Helmi2018,Bland-Hawthorn2019,Belokurov2019}.

Starting with the discovery of the Galactic HI spiral arms~\citep[][and references therein]{OortJ.H.;KerrF.J.;Westerhout1958}, it has been clear that the Sun is a star in a spiral galaxy. In the inner region, the Milky Way seems to be dominated by a massive bar \citep[e.g.,][]{Dwek1995} and an X-shaped structure \citep[e.g.,][]{Wegg2013}, similar to what is seen in extragalactic edge-on boxy bulges. As these are the most prominent dynamic features in the inner Galaxy, research of the kinematics and stellar populations that constitute the bar and the bulge is necessary to understand the morphology, structure and evolution of the Milky Way~\citep{Bland-Hawthorn2016}. Evolved stars, that are prominent in the mid-Infrared (mid-IR), are possibly the best targets for such studies~\citep{Kunder2012}. Indeed, the bar and bulge have been probed by counting IR stellar densities~\citep{Blitz1991, Babusiaux2005,Rich2007}, studying their metallicities and sometimes their variability, which for some stars can be used to obtain distance estimates.

Typically, these stars are too distant to measure proper motions or direct parallax distances from their stellar photosphere, as their Spectral Energy Distributions (SED) peak in IR, while their optical images are hidden behind circumstellar and interstellar dust. However, the most extreme of these evolved stars harbor circumstellar masers~\citep[see e.g.,][]{Hofner2018}. Circumstellar masers are useful as they are bright beacons of a specific evolutionary stage in which evolved stars develop a thick circumstellar shell with specific molecular content and exceptional physical conditions. Moreover, the masers deliver accurate line of sight velocities through the Doppler effect. Finally, stellar maser emission reaches high brightness temperatures, allowing in principle Very Long Baseline Interferometry (VLBI) astrometry with micro-arcsecond accuracy~\citep{vanLangevelde2003,Reid2014}.

Previous surveys focused first on OH masers~\citep{Sevenster2000,Fish2006} and later targeted SiO masers with single dish telescopes~\citep{Messineo2018}. When it was realized that the new capabilities at 7mm of the NSF's Karl G. Jansky Very Large Array (VLA) and 3mm of the Atacama Large Millimeter/submillimeter Array (ALMA) offer efficient ways to study SiO masers, the Bulge Asymmetries and Dynamical Evolution project (BAaDE\footnote{\url{http://www.phys.unm.edu/~baade/}}) was proposed. Using Midcourse Space Experiment (MSX) IR color selections, many thousands of SiO masers are found~\citep{Sjouwerman2017,Stroh}. This sample may thus facilitate a detailed study the kinematics of the bulge, bar and inner Galaxy.

Since only very few SiO masers are known from young stars~\citep{Colom2015}, those stars that show emission at 43 and/or 86 GHz are almost exclusively Asymptotic Giant Branch (AGB) stars. But stars of a very wide mass range are expected to spend time in this phase, as they become unstable towards the end of their lives. As a consequence, the ages of these star can vary considerably, ranging from 100 Myr to a fraction of the age of the universe~\citep[e.g.,][and the references therein]{Salaris2014}. Metallicity effects also affect the observables of the AGB population, as stars for which the envelope becomes low in oxygen may not easily produce sufficient SiO~\citep[e.g.,][]{Sande2018}. Although the~{\gaia} mission cannot provide information on all of the BAaDE targets --- and certainly not the majority of targets that sample the inner Galaxy --- it can be used to characterize the stars in the BAaDE sample, particularly those in the local region ($d<$2 kpc). In this region, recent studies have detected major Galactic structures~\citep{Reid2019a,Alves2020} as well as several co-moving groups and stellar structures~\citep{Kounkel2020}.

In this paper, we cross-match the BAaDE sample with 2MASS and~{\gaia} DR2. Because the BAaDE sample is based on MSX, it predominantly contains stars at low Galactic latitude. The cross section of the various surveys has IR as well as optical astrometric information. Through the~{\gaia} DR2, we can evaluate other parameters such as the parallax and proper motion, but also information derived from the survey such as variability and stellar classification. The objective of this work is to understand the nature of stars that enter the BAaDE survey. As we selected objects from their $IR$ colors in MSX~\citep[with SiO maser emission detected for $\sim$$70\%$,][]{Trapp2018}, one can expect it to contain predominantly Long Period Variable (LPV) stars, likely Miras, with a modest circumstellar shell. But this sample may contain Young Stellar Objects (YSOs), Main Sequence (MS) or Red Giant Branch (RGB) stars, that are very luminous~\citep{Lewis2020}, or older, less massive stars that progress on the AGB track with lower luminosity. In order to address these issues, we present the cross-matches in Section~\ref{sec_observations}. In the following sections, we present the main features of the different samples that resulted from the cross-matching. We start with the {\it BAaDE-Gaia} cross-match sample described in Section~\ref{sec_baade_gaia}. Afterwards, we filter the sample to just the solar neighborhood defining the~{\local} sample in Section~\ref{sec_filtering}. In Section~\ref{sec_local}, we characterize the~{\local} sample in terms of luminosity, variability, Galactic distribution and Period-Luminosity relation. These proprieties position us to comment on the nature of evolved stars in the foreground Galactic plane, for which we have~{\gaia} counterparts with accurate distances. In a subsequent paper (Quiroga-Nunez et al. in prep), we will present an extended study of the features (e.g., kinematics, SiO maser emission and rates, carbon and oxygen rich stellar discernment, etc.) of the evolved stellar population in the Galactic foreground by using the current results of the BAaDE survey. 

\section{Cross-match at different wavelengths}
\label{sec_observations}

We have cross-matched the MSX-based BAaDE target sample with 2MASS and {\it Gaia} DR2 (see Table~\ref{tab_surveys}), using the {\it Gaia} data archive interface\footnote{\url{http://gea.esac.esa.int/archive/}}. This sample was defined as the {\it BAaDE-Gaia} sample. In the following subsections, we describe how this process was implemented, starting from the BAaDE target selection, followed by the cross-match criteria.

\subsection{BAaDE target sample selection}
\label{sec_baade_selection}

The BAaDE target selection was based on IR photometry and designed to identify red giant stars with envelopes likely to harbor SiO maser emission. Starting from the IRAS two color-color diagram (2CD), \cite{veen1988} studied dust and gas envelopes of AGB stars. They pointed out that circumstellar shell properties of AGB stars appear in a sequence in the IRAS 2CD, suggesting an evolutionary track with increasing mass-loss rate. In the IRAS 2CD, SiO maser stars are expected to be found in a specific color region, facilitating a selection based on the IRAS colors. However, the
angular resolution of IRAS varied between about 0.5\arcmin~to 2\arcmin~limiting a large-scale survey, particularly in the Galactic plane. Later, \cite{Sjouwerman2009} were able to transform the IRAS 2CD sequence to colors in the mid-IR, using MSX data. By doing this, the positional accuracy of identified IR sources was improved to 2\arcsec~\citep{Price1995}, and a new sample of AGB stellar candidates with mid-IR information was obtained. This way, 28,062 stellar targets were selected with the objective to sample the evolved stellar population in the Galactic plane, bar and bulge, mostly limited to $|b| < 5$\adeg{}. It is expected that one third of the BAaDE target sample lies in the Galactic bulge~\citep{Sjouwerman2017}. The 28,062 targets are being followed up in order to detect SiO maser emission at 43 GHz with the VLA or 86 GHz with ALMA. So far, 20,600 candidates have been observed, of which 16,335 have already been analyzed (14,548 with the VLA and 1,787 with ALMA) and the scientific products are planned to be released publicly soon. The remaining sources are expected to be observed with ALMA in future cycles.

\newcommand*\luisas{\includegraphics[scale=0.04]{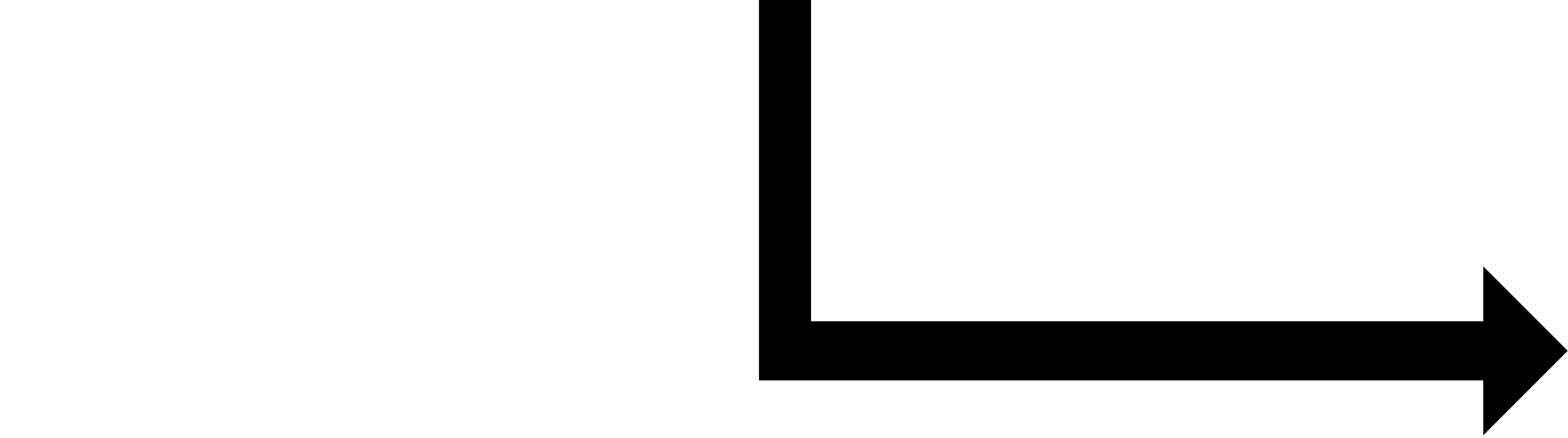}}

\begin{deluxetable}{lllllr}[htb]
\tablehead{
\colhead{Surveys} & \colhead{} & \colhead{} & \colhead{} & \colhead{} & \colhead{Sources}}
\startdata
BAaDE (MSX) & \colhead{} & \colhead{} & \colhead{} & \colhead{} & 28,062 \\
$\cap$ & 2MASS & \colhead{} & \colhead{} & \colhead{} & 25,809  \\
& $\cap$ & {\gaia} DR2 & \colhead{} & \colhead{} & 20,111  \\
\tablecaption{Sources numbers obtained for different samples and cross-matches.\label{tab_surveys}}
\enddata
\tablecomments{%See Section~\ref{sec_baade_selection} 
The intersection symbol ($\cap$) indicates cross-match between the surveys.}
\end{deluxetable}

\begin{figure}
\begin{center}
\resizebox{\hsize}{!}{\includegraphics{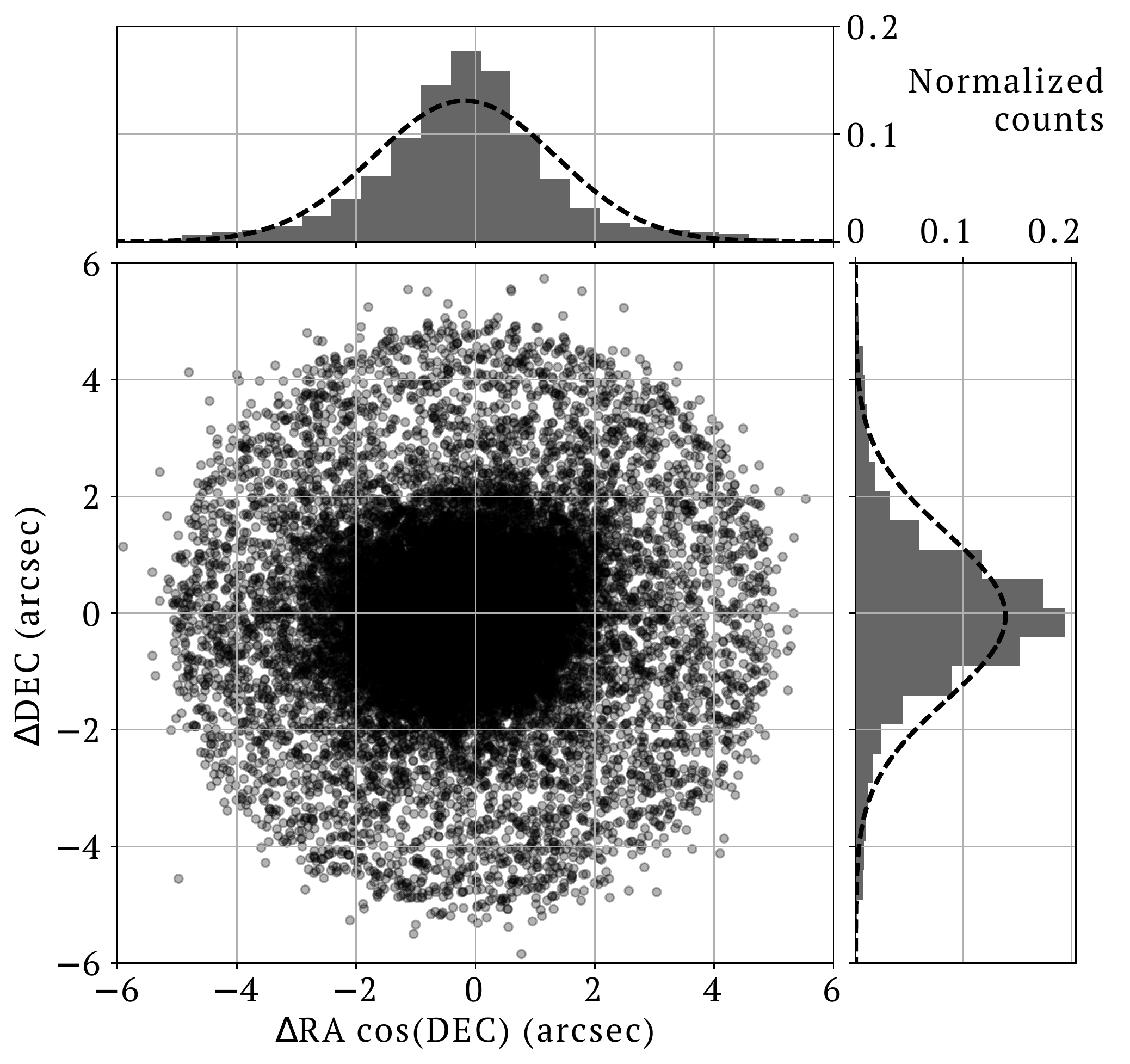}}
\caption{Distribution of the equatorial coordinate offsets between BAaDE targets and~{\gaia} DR2 counterparts. Each offset component can be well-represented by a 1D Gaussian distribution 
(see Section~\ref{sec_xmatch_description}).}
\label{fig_offset}
\end{center}
\end{figure}

\begin{figure*}
\begin{center}
\resizebox{\hsize}{!}{\includegraphics{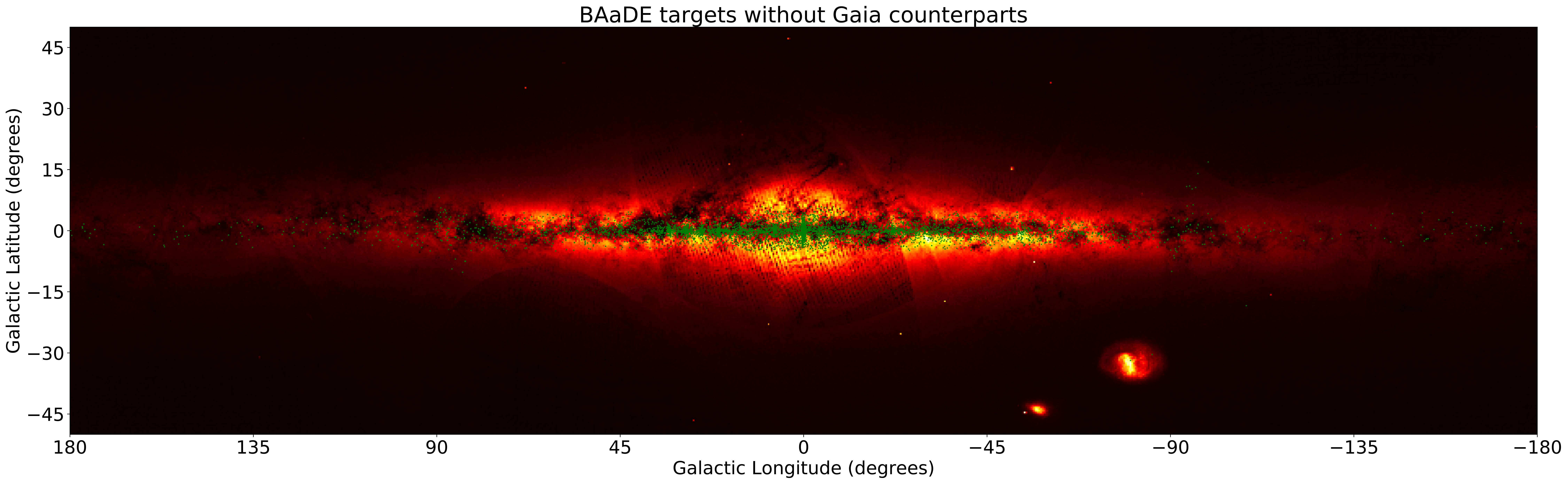}}
\caption{Galactic distribution of the BAaDE stellar targets without a~{\gaia} counterpart (green points) overplotted on the sky map from {\it Gaia} DR1. This sample accurately correlates with highly obscured regions in the optical regime. Credit: ESA/{\gaia}/DPAC.}
\label{fig_Gaia_nocounter}
\end{center}
\end{figure*}

\begin{figure*}
\begin{center}
\resizebox{0.85\hsize}{!}{\includegraphics{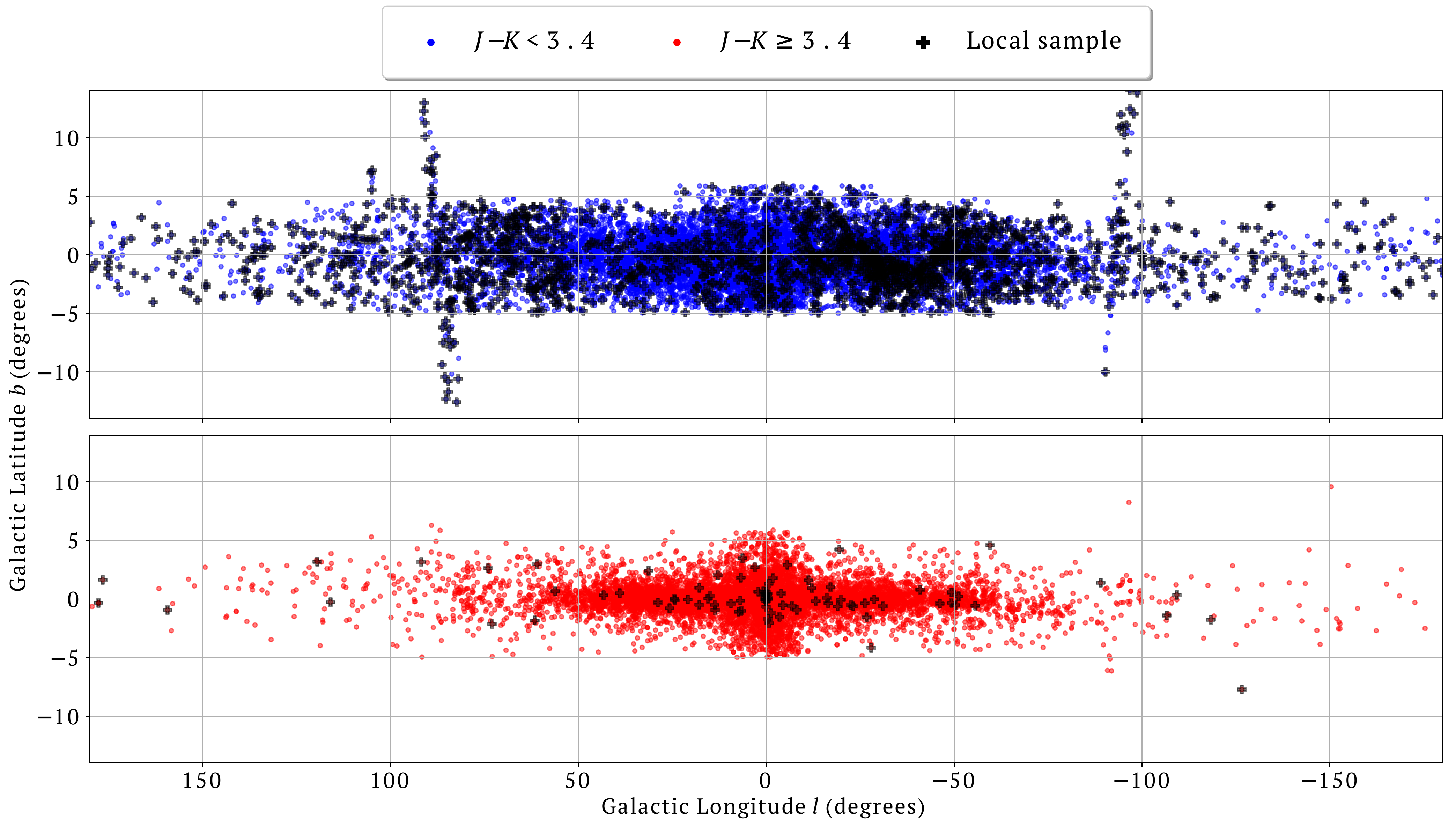}}
\caption{Galactic longitude-latitude diagram for the cross-matches obtained between BAaDE, 2MASS and {\it Gaia} defined as the {\it BAaDE-Gaia} sample. This sample was split in two populations (upper and lower panel) based on the mean 2MASS color $(J-K)$ obtained, similar to what~\cite{Trapp2018} have done to identify ``cold'' and ``hot'' kinematic populations using a subset of the BAaDE survey. Black crosses represent the defined~{\local} sample, a subsample of evolved stars in the {\it BAaDE-Gaia} sample with accurate parallax measurements at $<$2 kpc distance (see Section~\ref{sec_filtering} and~\ref{sec_local} for further details). The~{\local} sample is mainly made of foreground Galactic stars. The linear features observed at $l$$\sim$85\adeg{} and $l$$\sim$$-$85\adeg{} for sources with $|b|>8$\adeg{} are part of MSX target list (and are also BAaDE targets) caused by the target selection made by MSX in order to fill the
strips missed by the all-sky survey by IRAS~\citep{Egan2003}.}
\label{fig_l_b}
\end{center}
\end{figure*}

\begin{figure*}
\begin{center}
\resizebox{0.65\hsize}{!}{\includegraphics{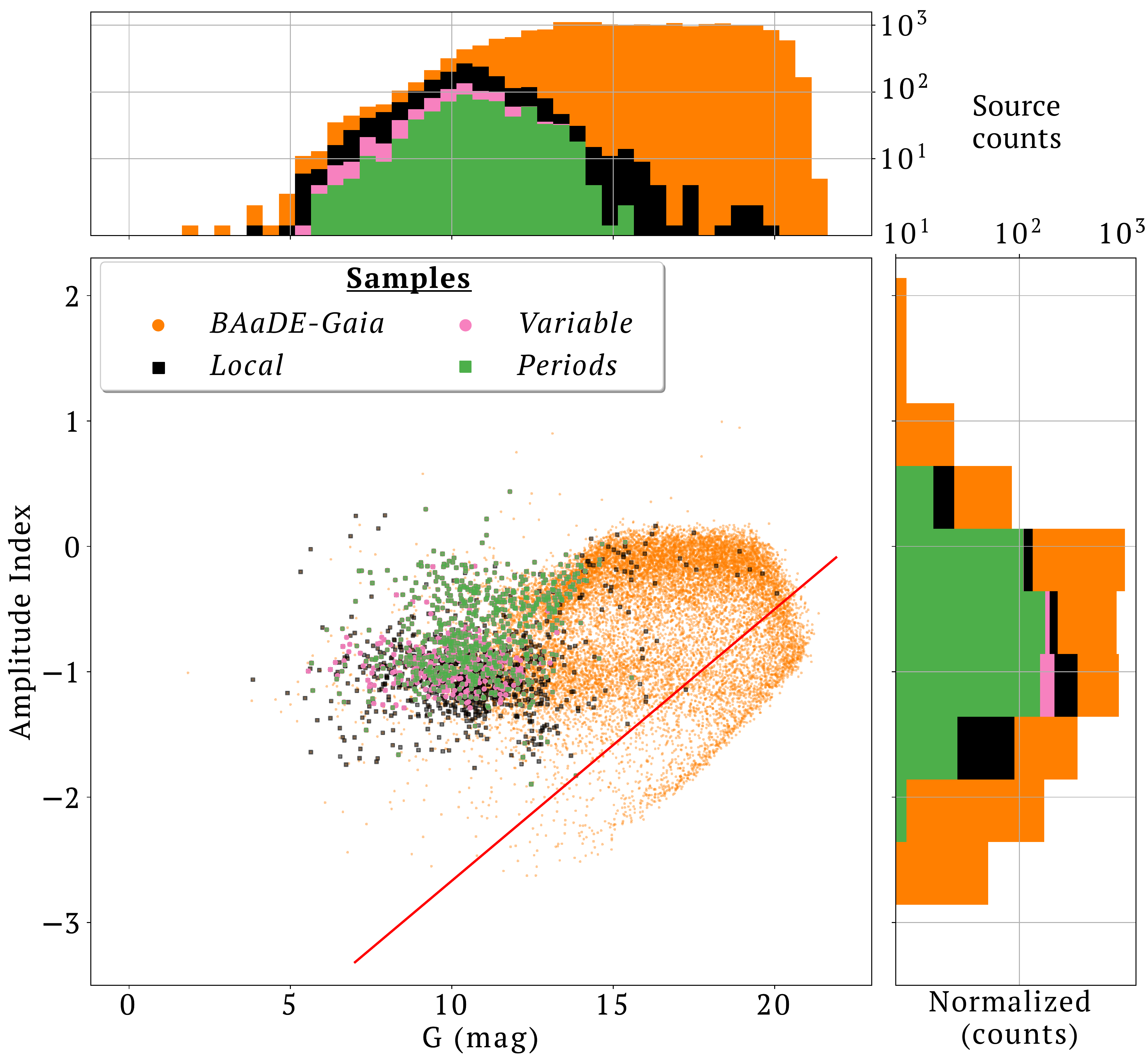}}
\caption{Amplitude-magnitude diagram firstly suggested by~\cite{Belokurov2017}, 
used to distinguish variable stars where larger amplitudes are likely associated with pulsating AGB stars. The orange points represent the {\it BAaDE-Gaia} sample (BAaDE $\cap$ 2MASS $\cap$ {\it Gaia} DR2), whereas black sources represent the stars within the~{\local} sample (i.e., accurate distance estimates for evolved stars at$<$2 kpc from the Sun). The pink and green points show two subsamples (\textit{Variable} and \textit{Periods}) derived from the~{\local} sample (see Table~\ref{tab_filters} and Section~\ref{sec_local}). The solid red line represents a threshold above which sources are predicted to show variability
larger than expected for a constant star at the given $G$ magnitude~\citep{Belokurov2017}. Possible features present in the plot (e.g., the apparent bimodality of the amplitude index) will be further explored in a subsequent paper}.
\label{fig_mag_g}
\end{center}
\end{figure*}

\subsection{Cross-match description}
\label{sec_xmatch_description}

In order to match the BAaDE targets with other surveys in position, we considered a conservative sky-projected circular area with 3\arcsec~radius around the BAaDE targets. The motivation for this separation was based on the assumption that the distribution of deviations from the actual positions is dominated by the MSX data~\citep[as confirmed by][]{Pihlstrom2018a} and has Gaussian distributions in both components $\rm{(\Delta \alpha \ cos(\delta), \Delta \delta)}$ with absolute mean values $<0.2$\arcsec~and positional accuracy of around 2\arcsec, as seen in Figure~\ref{fig_offset}. Therefore, a 3\arcsec~radius was chosen as a conservative match radius between the BAaDE targets and~{\gaia}. 
Note that the criterion we used here is more restrictive than the first cross-match done for a pilot of BAaDE sources and 2MASS~\citep[i.e., 5\arcsec~in][]{Trapp2018}.

After defining a sky-projected circular area, we proceed with cross-matching the BAaDE target sample with 2MASS and {\it Gaia} DR2. Although the cross-match in principle can be done directly with~{\gaia}, as it has typically lower positional offsets with respect to SiO masers positions~\citep{Pihlstrom2018a}, we started instead by cross-matching BAaDE and 2MASS, motivated by two different arguments. First, BAaDE targets that display both mid-IR emission (MSX) and optical emission ({\gaia}) are not expected to be extinct at NIR (2MASS). Hence, by initially cross-matching with 2MASS, we are already avoiding some false positives. Second, the cross-match between 2MASS and~{\gaia} was already established by~\cite{Marrese2018}, using a robust best neighbor algorithm, which found more than 90$\%$ overlap between both surveys. 

By using the 2MASS survey, we have found 25,809 counterparts for the BAaDE target list. Next, after cross-matching with {\it Gaia} DR2, the sample was reduced to 20,111 cross-matches (see Table~\ref{tab_surveys}), where all of them were found to be one-one correspondences. This last sample of 20,111 is called {\it BAaDE-Gaia} sample throughout the paper and thus includes 2MASS information. Notably, for 7,951 BAaDE targets (33$\%$), there were no {\it Gaia} counterparts, probably due to the fact that these targets lie behind considerable dust extinction at optical wavelengths. Figure~\ref{fig_Gaia_nocounter} shows how the distribution of these ``missing'' sources indeed correlates with the dust obscured regions that {\it Gaia} could not penetrate. 

%\vspace{0.5cm}

\subsubsection{Statistics of the cross-matches}

Assuming a uniform distribution of sources in the bulge for the~{\gaia} detections, as well as for BAaDE targets, one can calculate the number of sources that will give random matches at the given resolution of each survey. We estimated that the number of random matches should be less than 1,200; this is a small fraction of the 20,111 cross-matches that we have. Moreover, in this statistical estimate we have assumed that there is no optical extinction limiting the number of~{\gaia} sources. Therefore, the actual number of chance matches will be much lower than 1,200 indicating that our sample has at most a modest contamination of sources with unrelated counterparts.

\begin{deluxetable*}{lrlrlrr}[htb]
\tablehead{
\colhead{Sample name} & \colhead{Description or filters used} & \colhead{} & \colhead{} & \colhead{} & \colhead{} & \colhead{Sources}}
\startdata
{\it BAaDE-Gaia} & BAaDE (MSX) $\cap$ 2MASS $\cap$ {\it Gaia} DR2 & \colhead{} & \colhead{} & \colhead{} & \colhead{} & 20,111\\
%\colhead{} &\luisas & Variable & \colhead{} & \colhead{} & \colhead{} & 3,017\\
\colhead{} &\luisas & $\sigma_{\varpi}/\varpi<0.2$ &\colhead{}&\colhead{}&\colhead{}& 2,277\\
{\it Local}&\colhead{}& \luisas& $r<$2 kpc $\&$ evolutionary stage & \colhead{} & \colhead{}&1,812\\
{\it Variable}& \colhead{} & \colhead{} & \luisas & Variable  & \colhead{} & 898\\
{\it Periods}& \colhead{} & \colhead{} & \colhead{} & \luisas & Periods& 649
\tablecaption{Number of sources obtained for each subsample of the {\it BAaDE-Gaia} sample. Each row represents a filter used. See Section~\ref{sec_local} for a detailed description of each filter. The definition of the {\it BAaDE-Gaia} and {\it local} samples is given in Sects.~\ref{sec_xmatch_description} and~\ref{sec_local}, respectively. \label{tab_filters}}
\enddata
\tablecomments{The arrow symbols indicate subsample, where as the intersection symbol ($\cap$) indicates cross-match between the surveys.}
\end{deluxetable*}

%\vspace{2mm}
\section{Features of the {\it BAaDE-Gaia} sample}
\label{sec_baade_gaia}

Since the {\it BAaDE-Gaia} sample was obtained through 2MASS, the mean value of near-IR color ($J-K$) can be used to split the sample in two equal sized subsamples: i.e., $(J-K)<3.4$ for the bluer stars and $(J-K)\geq3.4$ for the redder stars. More extreme AGB stars (more luminous and with thicker shells) are expected to have steeper slopes in their SEDs at near-IR wavelengths, resulting in increasingly redder IR colors. Figure~\ref{fig_l_b} shows the subsamples of red and blue stars in a Galactic latitude-longitude diagram. Red stars seem to better trace the inner part of the Galaxy (Galactic bulge and plane) while bluer stars seem to dominate the foreground population. Indeed, as we will detail in Section~\ref{sec_filtering} and Section~\ref{sec_local}, Figure~\ref{fig_l_b} also shows that sources in the solar neighborhood ($<$ 2 kpc) are mainly stars that are bluer (in the context of the BAaDE selection) in particular those with~{\gaia} counterparts. We confirm that by splitting the sample using IR photometry, two samples can be traced. This has already been observed by~\cite{Trapp2018}, who made the split using $K$ magnitudes, and labeled the two a kinematic populations ``cold'' (the bluer, brighter stars in the Galactic disk) and ``hot'' (the redder stars in the bulge/bar). However, although it is indeed expected that more extreme stars are redder, we must highlight that the increased extinction with distance (toward the bulge) also makes them redder. Therefore, this partly explains why these stars show up nicely as bulge sources (Figure~\ref{fig_l_b}) and seem to be a better tracer of the inner galaxy.

Another property that can be investigated for the {\it Gaia-BAaDE} sample is variability. Although the {\it Gaia} DR2 has variability information for a considerable number of stars~\citep{Mowlavi2018a}, \cite{Belokurov2017} has shown that ---already with {\it Gaia} DR1--- flux uncertainties quoted in the~{\gaia} catalogue reflect the dispersion of the $G$-band flux measurements, which will thus lead to apparently larger uncertainties for variable stars. They have defined an amplitude variation over error, which we refer as amplitude index throughout this paper, using the mean flux ($\overline{I_g}$) and its error ($\sigma_{\overline{I_g}}$) in the optical $G$-band as $\rm{log_{10} \left ( \sqrt{N_{obs}}\frac{\sigma_{\overline{I_g}}}{\overline{I_g}}\right )}$, where $N_{obs}$ is the number of observations. Using this quantity, \cite{Belokurov2017} calculated the amplitude for different stellar populations in the Large Magellanic Cloud (LMC), finding that Mira variables have an amplitude index $>-1.0$. Figure~\ref{fig_mag_g} shows an amplitude-magnitude plot for the~\textit{Gaia-BAaDE} sample, where stars with amplitudes larger than $-1$ in this diagram are likely pulsating stars. However, although the amplitude index seem to be an useful tool to estimate variability, it might have issues when it is applied to a sample different in properties from the Magellanic system, or when more observations become available ($N_{obs}$). Therefore, we could expect that the cut-off changes depending on the stellar population.

Figure~\ref{fig_mag_g} also shows that the variable stars defined by~\cite{Mowlavi2018a} within~{\gaia} DR2 (see Section~\ref{sec_local}) coincide with larger amplitude values as expected, confirming that indeed the IR classification made by the BAaDE project correlates with variable stars. However, this qualification is restricted to stars that are bright in the $G$ band.

\section{Filtering the {\it BAaDE-Gaia} sample for Galactic foreground sources}\label{sec_filtering}

As the objective of this study is to characterize the evolved stars in the BAaDE target list, we apply additional refinements of the cross-matches in order to identify contaminating sources. Several filters have been considered, which in turn have generated several subsamples from the {\it BAaDE-Gaia} sample of 20,111 sources. Below, we outline the criteria that have been applied, finally arriving at the resulting sample of evolved stars in the foreground Galactic plane, which we define as the~{\local} sample. Table~\ref{tab_filters} summarizes the resulting subsamples.
\subsection{Parallax measurements}
\label{subsec_plx}

Obtaining distance estimates from noisy parallax measurements can be a complex issue~\citep[see e.g.,][]{Bailer-Jones2015}. Several tools are available to extract statistically robust distances from parallax measurements with limited accuracy --- even from negative parallaxes ---~\citep[see e.g.][]{Bailer-Jones2018,Luri2018}. However, such distance estimates strongly rely on robust expectations of stellar properties for a target sample. In our case, the best approach would be to compute the parameters of a probability distribution specifically for AGB stars by maximizing a likelihood function, so that under an assumed statistical model the distance distribution for the observed evolved stellar data is the most probable. However, if for a specific star $\sigma_{\varpi}/\varpi < 0.2$, one could obtain an accurate estimate of the distance without further considerations~\citep{Bailer-Jones2015}. In this sense, we find that most (91$\%$) of the stars of the~\textit{BAaDE-Gaia} sample that have a $\sigma_{\varpi}/\varpi < 0.2 $ are limited up to 2 kpc. Moreover, since (1) the aim of this research is to study the foreground population of evolved stars and (2) accurate extinction maps are limited to 2 kpc (see following subsection), we focus on the solar neighborhood ($<$2 kpc). Finally, an analysis of the foreground sample can be considered an initial step for doing a full statistical analysis. 

We have also investigated the effect of the {\it Gaia} parallax zero-point for our targets. In principle, the {\it Gaia} parallax zero-point can be up to 100 $\mu$as depending on the method and sample used~\citep[e.g.][and references therein]{Zinn2019,Chan2020}. However, several studies coincide that the {\it Gaia} parallax zero-point for red clump and variable stars oscillates around $-$50$\mu$as~\citep{Groenewegen2018a,Riess2018,Langevelde2018,Zinn2019,Chan2020}. Such an offset may cause a shift of less than 5$\%$ of the average values of the physical quantities reported in this work.

\subsection{Extinction maps up to 2 kpc}
\label{sec_extinction}

\cite{Capitanio2017} and \cite{Lallement2019} have produced local dust maps, based mainly on a regularized Bayesian inversion of individual color excess measurements using {\it Gaia} data. Additionally, the authors combined several tracers to confirm accurate extinction maps and reddening estimates up to 2 kpc. This tool is extremely useful to estimate intrinsic luminosities for the stars in our sample, which is an important physical property that can be used to characterize the stellar population. Although for local AGB stars, which emit mostly in the (mid-) IR, the effects will be small, we do adopt these maps, and thus a distance limit of 2.0 kpc.

\subsection{Younger stars detected in HR diagrams}\label{sec_ysos}

\begin{figure}
\begin{center}
\resizebox{\hsize}{!}{\includegraphics{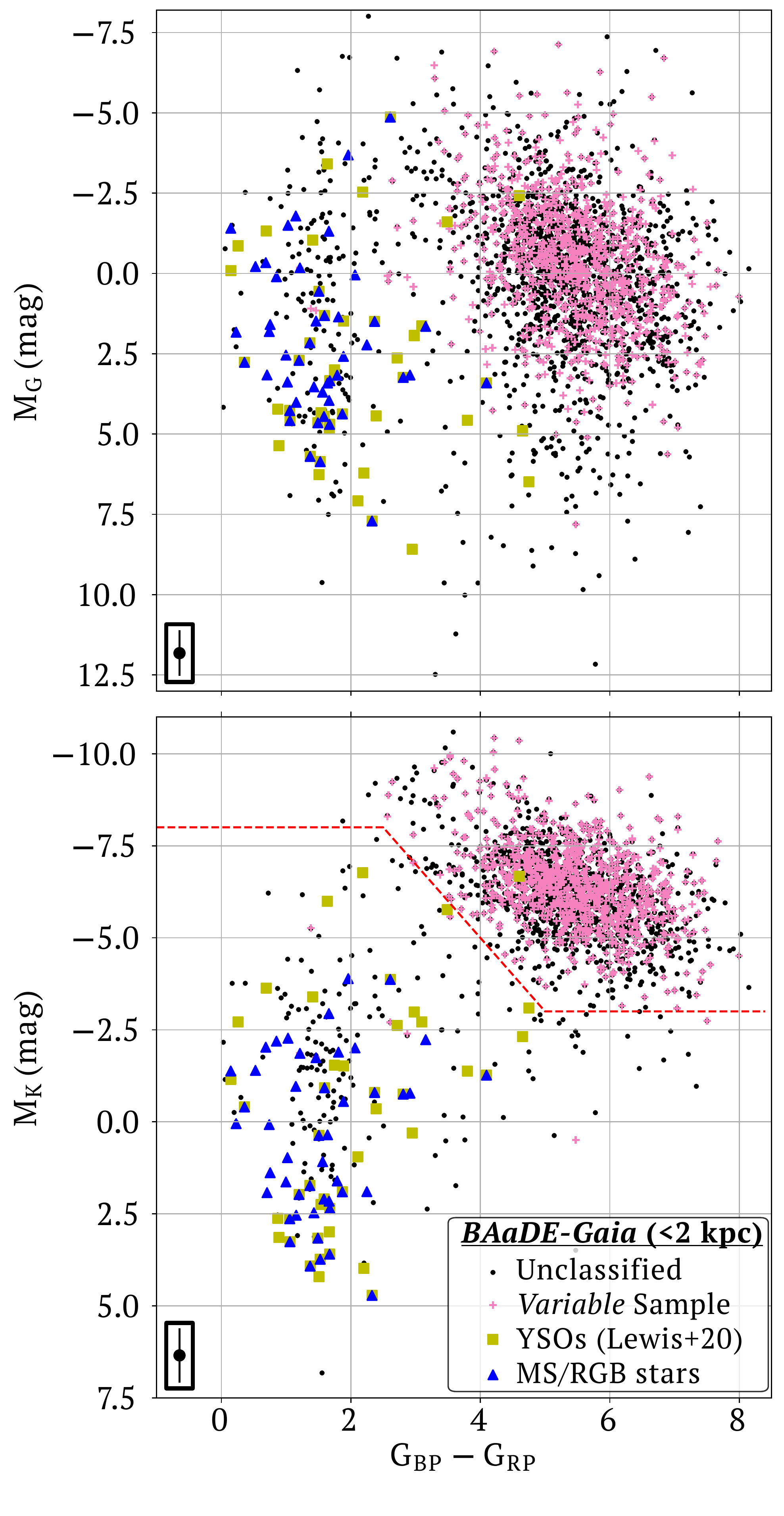}}
\caption{Absolute magnitude in G (\textbf{upper panel}) and $K$ (\textbf{lower panel}) bands as function of~{\gaia} color for the \textit{BAaDE-Gaia} sample with a distance $<$ 2 kpc. The typical errors in magnitude are shown in the left corner of each diagram. The sources in the sample that are also classified as MIRA (\textit{variable} sample, see Section~\ref{sec_local}) by~{\gaia} are marked as pink crosses. As yellow squares, we marked the sources that~\cite{Lewis2020} have confirmed as YSOs, where blue triangles are those likely associated with MS/RGB stars (Pihlstr\"om et al. in prep.). The dashed red line was established to distinguish evolved stars in the upper region (which defines the~{\local} sample) from other BAaDE targets in the lower region as YSOs, and MS/RGB stars (see Section~\ref{sec_ysos}). Absolute magnitudes in the $K$ band for typical low-metallicity Mira variables reported in the LMC by~\cite{Whitelock2008} cover a range between $-$8.0 and $-$6.0 magnitudes, whereas the AGB bump for~{\gaia} data seems to be located at $\rm{M_{G}\sim-0.5}$~\citep{Babusiaux18}.}
\label{fig_HRDs}
\end{center}
\end{figure}

Using the~{\gaia} DR2 results, \cite{Babusiaux18}
have generated several observational Hertzsprung-Russell diagram (HRDs). Particularly, for a sample of 29,288 low-extinction nearby giants (i.e., $>$ 2 mas parallax, E(B - V) $<$ 0.015 and $\rm{M_G} <$ 2.5), they were able to locate the AGB bump (at $\rm{M_G}$$\sim$$-$0.5 and $\rm{G_{BP}-G_{RP}}$$\sim$1.5), which corresponds to the starting point of the AGB where stars are burning their helium shell~\citep[see Figure 10 in][]{Babusiaux18}. In order to compare these results with the BAaDE targets in the foreground Galactic plane, we generate the HRDs shown in Figure~\ref{fig_HRDs}. These diagrams use~{\gaia} colors, $K$-apparent magnitudes from 2MASS, accurate distance estimates and extinction maps for the resulting~\textit{BAaDE-Gaia} cross-match around the Sun ($<$2 kpc). Moreover, the \textit{variable} sample was also over-plotted (pink crosses in Figure~\ref{fig_HRDs}) to support the statement that sources already classified as Mira by~{\gaia} fall in a defined location within the diagram (see Section~\ref{sec_local} for the definition of the \textit{variable} sample). This location is close to where the expected AGB bump is happening~\citep[$\rm{M_G}$$\sim$ 0.5,][]{Babusiaux18}, but with expected redder colors and fainter sources due to dust --- a combination of circumstellar and ISM reddening --- than those reported in the~{\gaia} HRD, as their sample was filtered for brighter sources sources~\citep[$\rm{M_G} <$ 2.5,][]{Babusiaux18}. Therefore, we conclude that sources concentrated around $\rm{M_G}$$\sim$$-$0.5 with redder colors are certainly AGB stars with significant circumstellar shells.

On the other hand, we have also confirmed that the IR color selection, by which the BAaDE targets were selected, does not prevent a fraction of YSOs and MS/RGB stars from entering in the target sample. In fact, \cite{Lewis2020} have shown that by using MSX colors, specifically the MSX color $[D] - [E]$, a differentiation between YSOs and AGB stars can be invoked within the BAaDE sources. Moreover, Pihlstr\"om et al. in prep., are identifying BAaDE sources that based on their IR photometry are likely associated with either reddened, massive MS stars (B or A stellar type) or RGB stars. These samples are also shown in Figure~\ref{fig_HRDs} to highlight their position in the HRDs, confirming that they are likely not AGBs. 

Since the distribution in $\rm{M_K}$ is less dispersed for the pre-selected AGBs sources, we made a cut in this diagram (lower panel in Figure~\ref{fig_HRDs}) to filter non-AGB sources. We define the~{\local} sample as the sources with the following~{\gaia} colors and absolute K-magnitude conditions:

$$
G_{BP}-G_{RP} \left\{
        \begin{array}{lll}
            <3 & \quad \rm{and }\quad \rm{M_k} < -8 \\
            >5 & \quad \rm{and} \quad \rm{M_k} < -3 \\
            \rm{rest} & \quad \rm{and} \quad \rm{M_k} \leq 2(G_{BP}-G_{RP})-13.            
        \end{array}
    \right.
$$

The non-AGB stellar sources were filtered out leaving 1,812 local stellar sources in the AGB regime. This means that we were able to confirm the evolved stellar evolutionary stage for most ($88\%$) of the BAaDE targets in the foreground Galactic plane. From those, there was only one source that~\cite{Lewis2020} classified as YSO. We have confirmed that this source falls very close to the empirical MSX color frontier defined~\cite{Lewis2020} for YSOs, and therefore, is likely a genuine AGB star after all.

\begin{figure*}
\begin{center}
\resizebox{0.8\hsize}{!}{\includegraphics{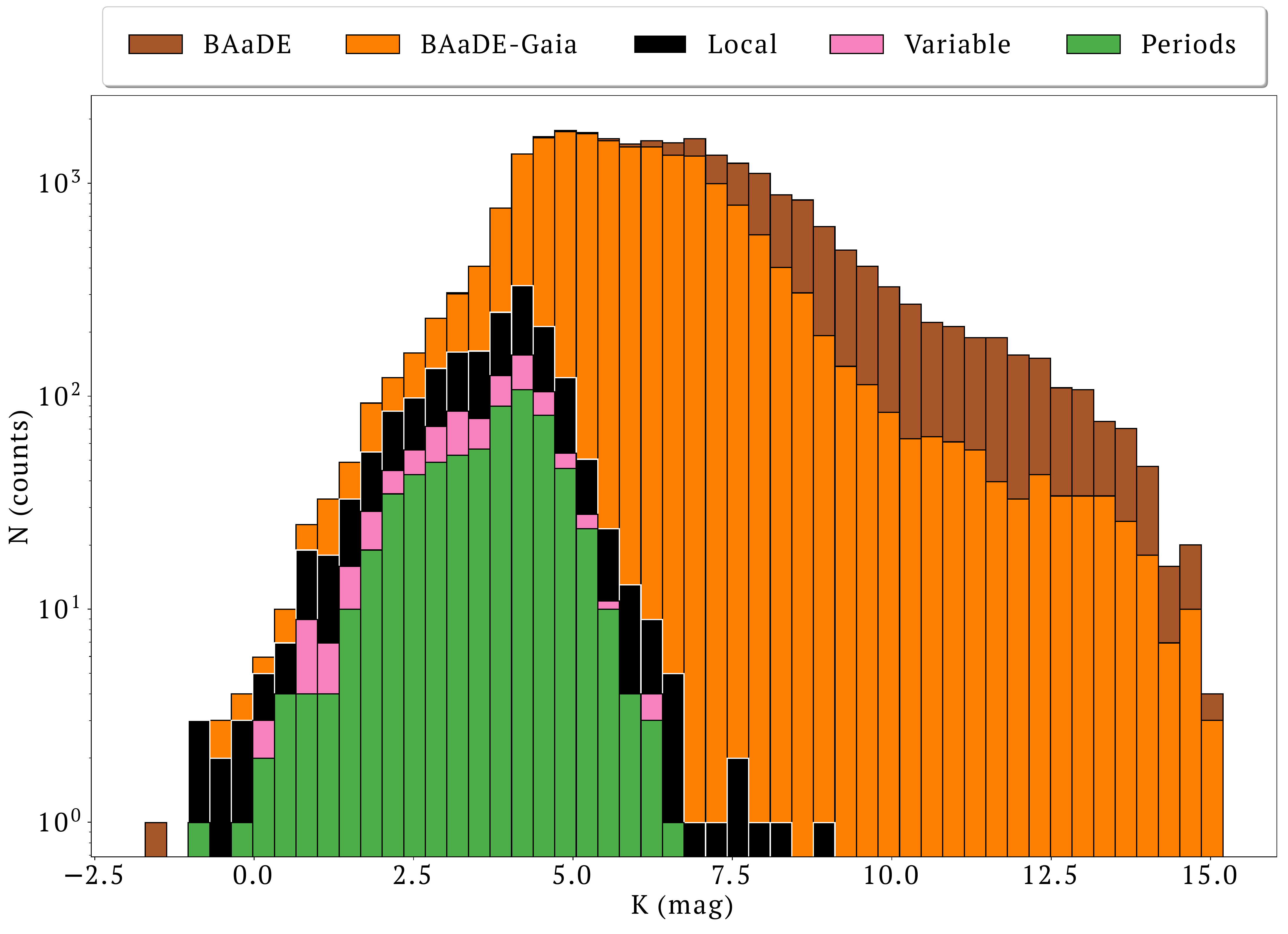}}
\caption{Histogram comparison for the distribution of the $K$-magnitude observed by 2MASS for different samples as shown in Tables~\ref{tab_surveys} and~\ref{tab_filters}.
The figure also shows that when more filters are applied, we are targeting the nearby stars, making the distribution narrower towards the apparently brighter stars.}

\label{fig_histo_mk}
\end{center}
\end{figure*}

%\vspace{5mm}
\section{The foreground population of evolved stars: the~{\local} sample}
\label{sec_local}

\begin{figure}
\begin{center}
%\resizebox{\hsize}{!}{\includegraphics{exctintion.png}}
\resizebox{0.9\hsize}{!}{\includegraphics{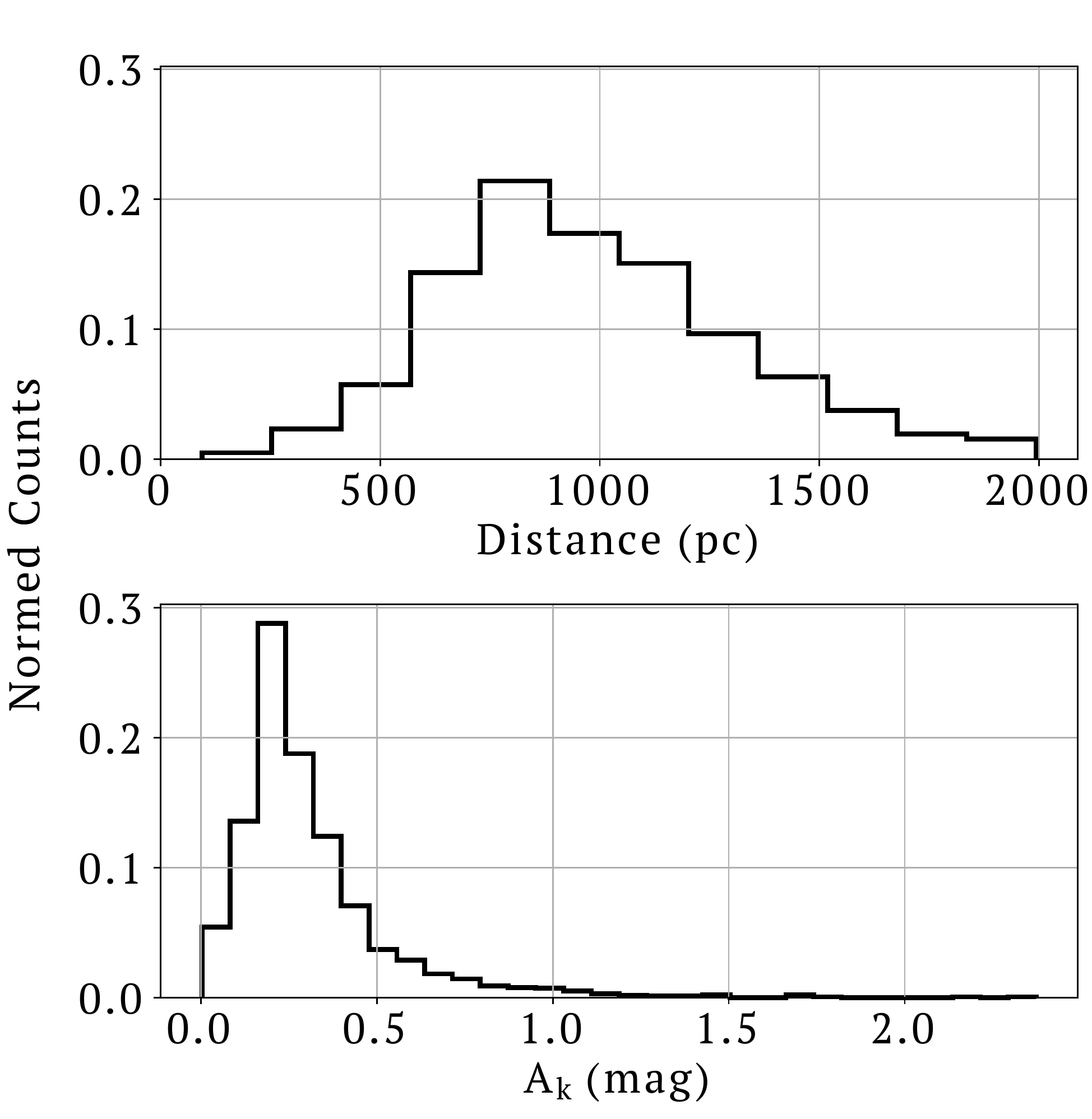}}
\caption{\textbf{Upper panel:} Distribution of distance to the Sun for the { \it local sample}. \textbf{Lower panel:} Extinction in the $K$-band obtained from the optical extinction maps developed by~\cite{Capitanio2017} and~\cite{Lallement2019}, and converted to IR $K$-band following~\cite{Messineo_thesis}.}
\label{fig_Ak}
\end{center}
\end{figure}

Using the 20,111 cross-matched sources that we have found between BAaDE, 2MASS and~{\gaia} DR2 ({\it BAaDE-Gaia} sample), we have applied the additional filters, previously described in Section~\ref{sec_filtering}, leaving a sample of 1,812 stellar sources that we have defined as~{\local} sample. This sample contains BAaDE targets associated with AGB stars within 2 kpc distance around the Sun with accurate distance estimates, IR and optical photometry and proper motions.

In addition, the~{\local} sample can be filtered by variability. For this, we have used the {\it Gaia} DR2 variability information contained in the~{\gaia} table {\tt vari\_classifier\_result}, and extracted those objects that were flagged as variables of any kind, which we define as the {\it variable} sample (898 sources). Next, we have refined the sample by extracting the sources with period estimates from the~{\gaia} table {\tt vari\_long\_period\_variable}, and named it as {\it periods} sample (649 sources). Note that all sources within the~{\local} sample contained in the table {\tt vari\_classifier\_result} were classified by~{\gaia} as Mira/semi-regular stars ({\tt MIRA\_SR}). The characteristics (variable and periods) described allowed us to generate subsamples of the {\it local} sample, as shown in Table~\ref{tab_filters}. 

Finally, it should be noted that there are two effects that play a role when distances to individual AGB stars are estimated. First, the strong colour variations of the stellar photosphere~\citep[see e.g.,][]{Lindegren2018,Langevelde2018}, and second, the photocenter movements caused by large atmospheres with convective motions~\citep{Chiavassa2018}. We have checked and added the {\it Gaia} {\tt astrometric\_excess\_noise} uncertainty when discussing individual objects.

In the following subsections, we research the Galactic foreground sample of BAaDE targets ({\local} sample) and its different subsamples (\textit{variable} and \textit{periods}) in terms of IR photometry, absolute and bolometric magnitudes, variability, Galactic distribution and the P-L relation. 

\subsection{Infrared photometry}

The SED of AGB stars usually peaks at IR wavelengths, therefore, these stars have been usually identified by their IR colors~\citep[see e.g.,][]{veen1988}. In particular, after the 2MASS data release~\citep{Skrutskie2006}, $K$ measurements have been widely used to characterize these populations~\citep{Whitelock2008,Messineo2018}. Figure~\ref{fig_histo_mk} shows the distribution of the apparent $K$ magnitude obtained from 2MASS for the entire cross-matched sample, with the different sub-samples in Table~\ref{tab_filters}. We note that by filtering the foreground sample with {\it Gaia} counterparts (the~{\local} sample), we are selecting brighter stars in the $K$-band. 

By using the optical extinction maps described in Section~\ref{sec_extinction}, we obtain the extinction and reddening estimates at $K$-band by assuming $\rm{A_{\lambda}/A_{\it k} = (\lambda/2.12 \ \mu m)^{-1.9}}$~\citep{Messineo_thesis}. As one could expect for the region around the Sun, the IR extinction estimates at the IR $K$-band for the filtered sample are usually lower than 0.5 mag (see lower panel of Figure~\ref{fig_Ak}). 
Finally, as we have accurate distance estimates for the~{\local} sample (see upper panel of Figure~\ref{fig_Ak}), we are able to estimate the absolute $K$ magnitude distribution ($\rm{M_K}$ in Figure~\ref{fig_Mk}) as is described in Section~\ref{sec_absolutemagk}.

\subsection{Absolute magnitudes for the foreground Mira population}
\label{sec_absolutemagk}

\begin{figure*}
\begin{center}
\resizebox{0.7\hsize}{!}{\includegraphics{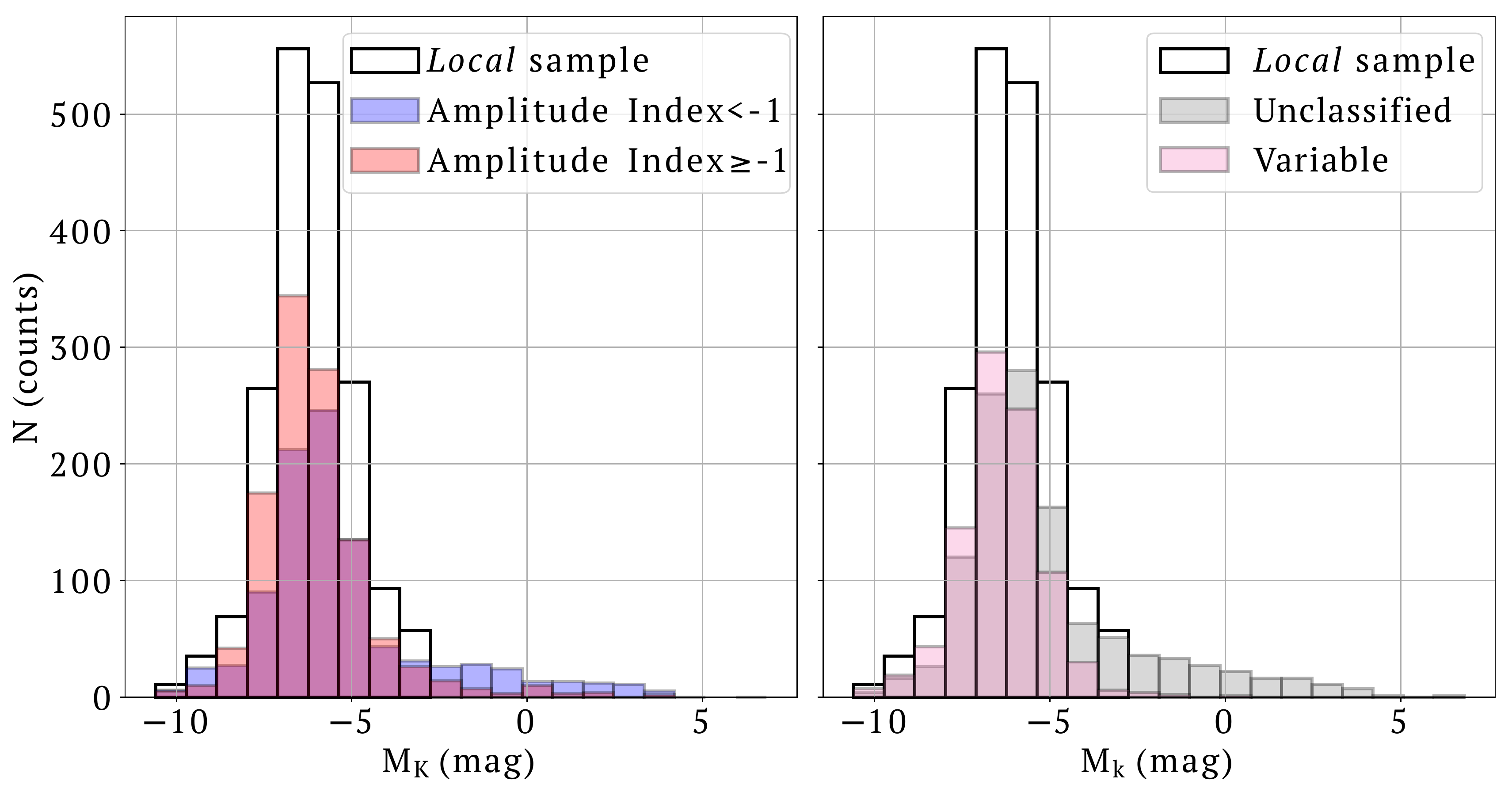}}
\caption{Each panel shows the absolute $K$-magnitude distribution as filled bars for the~{\local} sample before removing by YSOs and MS/RGB stars, and split by the variability criteria. These criteria are the Amplitude index~\citep{Belokurov2017} and the~{\gaia} DR2 variability classification~\citep{Mowlavi2018a}, in left and right panel, respectively. The~{\local} sample (without YSOs and MS/RGB stars) is also shown for comparison as a black unfilled histograms. We found that the Amplitude index method seems to find variability in objects that were not classified as variables in the~{\gaia} DR2.}
\label{fig_Mk}
\end{center}
\end{figure*}

\begin{figure}
\begin{center}
\resizebox{\hsize}{!}{\includegraphics{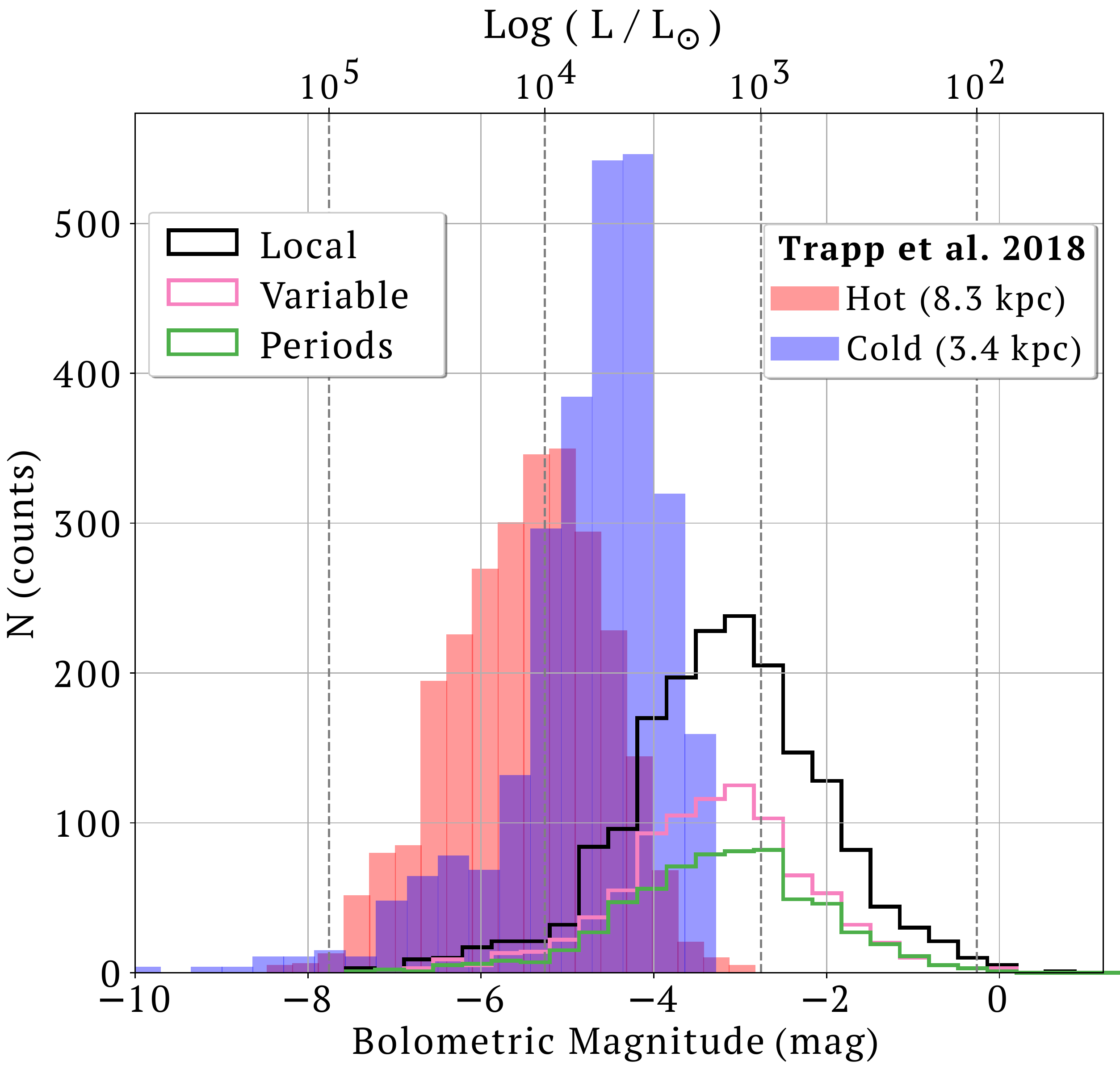}}
\caption{Luminosity and bolometric magnitude distributions for the~{\local} sample and the subsamples resulting from different filters (Table~\ref{tab_filters}). These distributions were obtained by applying the $BC$ proposed by~\cite{Messineo2018} to the absolute magnitudes in the $K$-band. The absolute $K$-magnitudes were estimated from 2MASS $K$-band, {\gaia} parallaxes and extinction maps from~\cite{Capitanio2017} and~\cite{Lallement2019}. The bolometric distribution estimated by~\cite{Trapp2018} for the ``hot'' and ``cold'' populations using an approximate kinematic model are shown as filled histograms. Note that the kinematic cold population proposed by~\cite{Trapp2018} is made up of stars in the Galactic disk and not in the bulge, therefore, similarities with respect to the~{\local} sample defined in this work are expected.}
\label{fig_bolometric}
\end{center}
\end{figure}

% \begin{figure*}
% \begin{center}
% \resizebox{0.75\hsize}{!}{\includegraphics{Presentation1.pdf}}
% \caption{\textbf{Conceptual map showing the variables and Bolometric Corrections (BCs) used to estimate the absolute $K$ magnitude ($\rm{M_K}$) in two different ways: $\rm{M_K}$ 2MASS and $\rm{M_K}$ {\gaia}-Messineo, in panels (a) and (b), respectively. The orange boxes represent different absolute magnitudes, blue boxes the stellar measurements and the green boxes the BCs used. We obtain the extinction and reddening estimates at $G$ and $K$-band by using optical dust maps~\citep{Capitanio2017,Lallement2019} and the relation proposed by~\citep{Messineo_thesis}, i.e., $\rm{A_{\lambda}/A_{\it k} = (\lambda/2.12 \ \mu m)^{-1.9}}$. The BCs shown for the $K$ and $G$ bands are based on~\cite{Messineo_thesis} and~\cite{Andrae2018}, respectively.}}
% \label{fig_conceptmap}
% \end{center}
% \end{figure*}

%AGB stellar populations have been characterized by IR absolute magnitudes, because they predominantly radiate at these wavelengths~\citep{Vassiliadis1993}. Such IR observations allows one to study large populations, given the low extinction at these wavelengths, but unbiased distance estimates can be hard to obtain.

Several studies have been carried out to estimate IR absolute magnitudes of the AGB populations in the LMC, where the distance to the stellar system is known, and therefore, the distance modulus (and presumably also the IR extinction) can be assumed the same for each object~\citep[see e.g., ][]{Whitelock2008}. From Figure~\ref{fig_HRDs}, we can determine an average absolute $K$-magnitude of $\rm{M_{k}=-6.3\pm1.2}$ mag for the~{\local} sample. Although the magnitude values found roughly correspond to those found in the LMC~\citep[i.e., between $-8.0$ and $-6.0$ mag, see e.g., ][]{Whitelock2008}, one should keep in mind that we established a fainter limit of $\rm{M_{k}=-2.5}$. Nevertheless, with the aim of analyzing the absolute magnitude distribution using the variability classification defined by~{\gaia}~\citep{Mowlavi2018a} and the amplitude index~\citep{Belokurov2017}, we made Figure~\ref{fig_Mk}, where the distribution of absolute magnitude for the {\it local} sample without considering the filtering of YSOs and MS/RGB (made in Section~\ref{sec_ysos}) is shown. Figure~\ref{fig_Mk} shows that by including these younger sources, the distributions are much broader in terms of absolute magnitude range. This can be partly explained, as the current sample is mostly based on a single 2MASS observation, and includes the effects of large amplitude variability. However, when we filter the sample by a variability qualification (as shown in both panels of Figure~\ref{fig_Mk}), the low luminosity tail (in $\rm{M_{k}}$) is cut out. In this sense, the~{\gaia} classification as Mira or Semi-Regular (SR) variable seems to narrow the distribution more. Following a similar discussion in \cite{Mowlavi2018a}, we argue that the low luminosity tail in Figure~\ref{fig_Mk} and also Figure~\ref{fig_histo_mk} is due to contamination with YSOs and MS/RGB stars that can also peak in the IR, but do not show the same variability~\citep[][Pihlstr\"om et al. in prep.]{Lewis2020}.

%\vspace{3mm}
\subsection{Bolometric magnitudes for the foreground Mira population}
\label{sec_bolo}

The bolometric luminosity is a fundamental property useful for classifying stellar populations and evolutionary stages~\citep{Srinivasan2009}, since it measures the intrinsic stellar power. Although its definition is straightforwardly formulated as the total integrated power over all frequencies, in practice, complete photometric measurements that allow a direct bolometric luminosity estimate are hardly ever available. Therefore, under various assumptions a limited set of photometric measurements, preferably near the peak of the SED, can be used to apply a bolometric correction ($BC$) in order to determine the integrated stellar luminosity.
In particular, for AGB stars IR absolute magnitudes are converted to bolometric luminosities using a bolometric correction, which is usually parameterized using IR colors~\citep[see e.g.,][]{Whitelock2008,Messineo2018,Lebzelter2019}.

\begin{figure}
\begin{center}
\resizebox{0.95\hsize}{!}{\includegraphics{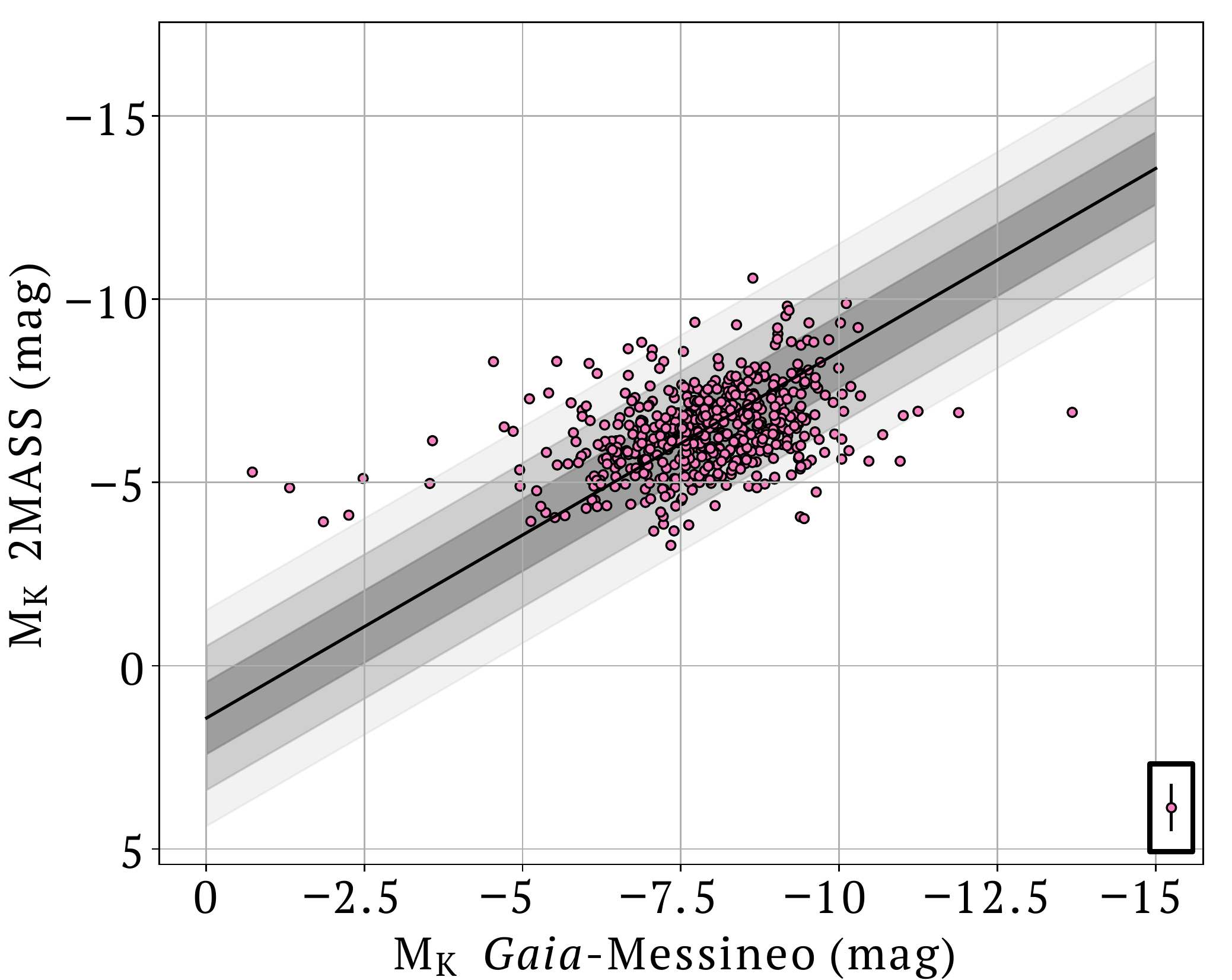}}
\caption{Comparison between the absolute $K$-magnitudes estimated from 2MASS data, {\gaia} parallaxes and extinction maps ($\rm{M_K}$ 2MASS, typical errors are shown in the right bottom corner) with respect to the absolute bolometric estimates reported for variable stars in~{\gaia} DR2 that were transformed to absolute $K$-magnitude using the $BC_{k}$ in~\cite{Messineo2018} ($\rm{M_K}$ \gaia-Messineo). The black line describes the linear fitting that was forced to have slope of 1. The grey layers contains 1, 2, and 3 $\sigma$ deviation from the linear fitting (see Section~\ref{sec_bolo}).}
\label{fig_MKvsMk_line}
\end{center}
\end{figure}

\cite{Trapp2018} have estimated the bolometric magnitude for a sub-set of the BAaDE sample. They considered a kinematically ``cold'' population of Galactic disk stars, which is similar to what is defined here as the foreground population or~{\local} sample. In their analysis, they have assumed a common distance of 3.8 kpc for this population and have applied a $BC_{k}$ based on~\cite{Messineo_thesis}. In order to compare the {\it local} sample with their kinematically ``cold'' population, we have applied the same $BC_{k}$, but not before confirming that other proposed $BC_{k_s}$ for AGB samples produced similar results~\citep{Whitelock2002,Srinivasan2009}. Figure~\ref{fig_bolometric} shows the bolometric distributions obtained. The offset between the bolometric distributions~\citep[the ones obtained for our samples, and those obtained by][]{Trapp2018} is likely caused by the distance assumption made by~\cite{Trapp2018}, which is equivalent to a range between 2.2 and 3.8 mag when taking the average distance of the~{\local} sample (1000 $\pm$ 350 pc, upper panel of Figure~\ref{fig_Ak}).

So far, we have based our bolometric magnitude estimates on $\rm{M_K}$ (Section~\ref{sec_absolutemagk}), which in turn was estimated by using 2MASS IR photometry, {\gaia} parallaxes, and extinction maps following the approach of~\cite{Messineo_thesis}. %This is sketched in the left panel of Figure~\ref{fig_conceptmap}, and 
We define it as $\rm{M_K}$ 2MASS. However, we can also obtain absolute $K$ magnitudes directly from~{\gaia} data, and we define it as $\rm{M_K}$~{\gaia}-Messineo. % (see right panel of Figure~\ref{fig_conceptmap}).
This one was calculated by using~{\gaia} photometry in the $G$ band, {\gaia} parallaxes and dust maps to estimate first the absolute $G$ magnitude for each star. Then, we use the $BC_{G}$ provided for~{\gaia} data~\citep{Andrae2018} to estimate the bolometric magnitude, and finally we estimate $\rm{M_K}$ using the $BC_{k}$ provided by~\cite{Messineo_thesis}. 

If we assume that both BCs produce similar results, then the $\rm{M_K}$ obtained for each star should be the same. In other words, a plot between $\rm{M_K}$ 2MASS and $\rm{M_K}$~{\gaia}-Messineo should produce a linear relation with a slope of one and an intersect of zero, which is not the case. Figure~\ref{fig_MKvsMk_line} shows this plot where we have done a linear fitting by forcing a slope of one ---and therefore we are assuming that both $\rm{M_K}$ estimates must be equal--- finding an offset of $-1.4\pm1.0$ mag ($\rm{M_K}$ 2MASS value at $\rm{M_K}$ Gaia-Messineo=0), indicating that the bolometric corrections in the {\it Gaia} DR2 seem to be overestimating the total luminosity calculated by~\cite{Andrae2018} of our very red, variable objects. Further research comparing different stellar populations is needed to refine the~{\gaia} BC at least for these red stars. We continue to use only the $\rm{M_K}$ based on 2MASS IR photometry, {\gaia} parallaxes and extinction maps, i.e., $\rm{M_K}$ 2MASS.

In Figure~\ref{fig_bolometric}, we also present the luminosity distribution for the {\it local} sample. It shows that our sample is made up of giant stars with a luminosity range that is consistent with AGB stars, mostly Mira variables~\citep{Srinivasan2009}. Compared to previous studies of Mira variables in the LMC or Galactic bulge (where fixed distances have been assumed), we have found less luminous objects. This of course is expected in our selection that was based on a combination of IR detections, optical {\it Gaia} counterparts, extinction maps and distance selection. We found that the typical luminosity for the {\it local} sample is estimated as $1,500^{+3000}_{-500} \ L_\odot$ suggesting that evolved stars in the solar neighborhood are found to be moderately luminous stars, likely associated with low-mass stars. 

Another plausible explanation of the moderately luminous stars found in the solar neighborhood could come from $BC$ used in this research. In principle, we can affirm that we are obtaining an accurate absolute $K$- and $G$-magnitude distribution for the~{\local} sample sources given that (1) the magnitude ranges found are expected for AGB stars, and (2) their evolutionary stages were confirmed by comparing with~{\gaia} HRDs (see Figure~\ref{fig_HRDs} and Section~\ref{sec_ysos}). Therefore, if there is a miss-calculation in the bolometric estimates, it could come from the $BC$ proposed by~\cite{Messineo_thesis}, as this $BC$ was determined for AGB stars located in the bulge. The metallicity difference with respect to nearby AGB stars, could cause a significant change in the luminosity estimate made. A planned research using BAaDE targets in the bulge would clarify this point.

\subsection{Variability}

At the beginning of Section~\ref{sec_local}, we have described the tables from {\it Gaia} DR2 that yield the variability classification that can be used for the {\it local} sample. Moreover, we noted that the ratio between the flux error and mean in $G$ magnitude can be used to identify pulsating stars when the amplitude index $>-1$. We have considered both methods, in particular in relation to the $K$-band apparent and absolute magnitude distributions. 
%Figures~\ref{fig_histo_mk} 
Figures~\ref{fig_mag_g} and~\ref{fig_Mk} show these distributions split according to both variability criteria. Although both methods seem to produce similar results, the variability criterion from the {\it Gaia} DR2 tables, achieves narrower ranges of absolute magnitudes (particularly for less luminous objects). In other words, the amplitude estimator based on the $G$ variance can presumably also pick up variability from objects that are not classified as variables in the {\it Gaia} DR2.

\subsection{Spatial distribution in the solar neighborhood}
\label{sec_sample2_gala}
\begin{figure}
\begin{center}
\resizebox{\hsize}{!}{\includegraphics{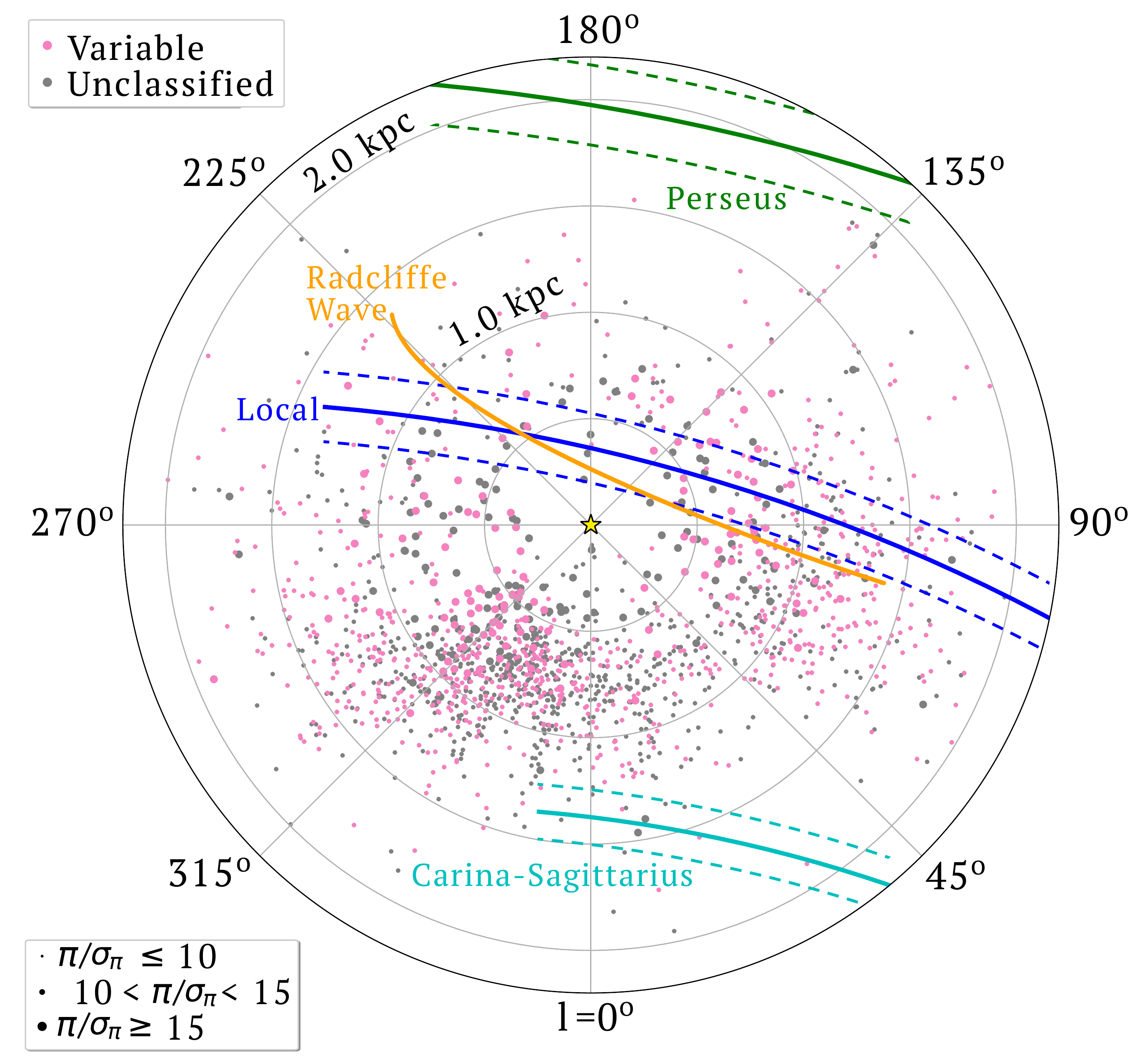}}
\caption{Foreground Galactic distribution of the~{\local} sample. The sample was split ``Variable'' and Unclassified according to~{\gaia} DR2~\citep{Mowlavi2018a}. The size of the marker is a measure of the relative parallax uncertainty, and therefore, distance uncertainty for each source. The positions and widths of the spiral arms are based on~\cite{Reid14}, whereas the Radcliffe wave is based on~\cite{Alves2020}. These structures were found not to be correlated with the occurrence of the evolved stellar sources (see Section ~\ref{sec_sample2_gala}). The stellar distribution found relies on (1) the BAaDE target selection made and (2) the dust distribution~\citep{Capitanio2017,Lallement2019}.}
\label{fig_Galactic}
\end{center}
\end{figure}

The Galactic distribution of AGB stars has been studied extensively using IRAS, WISE, 2MASS and MSX data~\citep{Jackson2002, Habing1996, Lian2014a,Messineo2018,Sjouwerman2009}. Generally, it has been found that AGB stars are tracing the dynamically relaxed stellar population of the Galactic thick disk. This is expected as these old stars have already migrated from their birth place being now detached from the spiral structure usually traced by young massive stellar objects~\citep{Quiroga-Nunez2017,Reid2019a}. In particular, \cite{Jackson2002} found a density distribution based on revised IR photometric data from IRAS that they called universal, implying that there are no statistically significant differences in the spatial distribution of AGB stars based on IR colors. Adopting their radial scale length of 1.6 kpc (outside of $R>5$ kpc) and scale height of 300 pc, we consider Figure~\ref{fig_Galactic}; which shows the projected spatial distribution of the~{\local} sample. 

We have found that the number density of sources in the solar vicinity ($<$0.5 kpc) is considerably lower than further out. The depletion of targets around the Sun originates from the fact that the MSX catalogue, on which our sample is based, is mostly limited to $|b|<$ 5\adeg{}. Therefore, the volume sampled increases with distance. From this, we estimate a corresponding scale height of $\sim$50 pc, equally for most of the objects in the {\it local} sample ($>$0.5 kpc). This seems to suggest that the scale height for our BAaDE targets is lower than scale height of the Galactic disk, i.e., $\sim$300 pc~\citep[][]{Jackson2002}.

Recent studies using hundreds of maser bearing stars have suggested a correlation between the position of evolved stars and the spiral arm structure at larger Galactic scales~\citep[up to 6 kpc,][]{Gorski2020,Urago2020}. In our study, however, we have found that this is not the case at least for AGBs in the solar neighborhood ($<$2 kpc). This can be seen in Figure~\ref{fig_Galactic}, where there is not a clear correlation with any of the two major Galactic structures in the region: the local spiral arm~\citep{Reid14}, and the recently discovered Radcliffe wave~\citep{Alves2020}. We must add that the {\it local} sample is affected by the interstellar extinction in the~{\gaia} bands, meaning that we might be biased to miss some sources at the highly extincted regions~\citep[i.e., the large Galactic structures as they are defined in terms of star-forming regions usually not reachable by~{\gaia},][]{Quiroga-Nunez2019}. Moreover, there is still a radial gradient detectable with more targets towards the center than observed in the anti-center direction. This arises due to the MSX criteria defined by \cite{Sjouwerman2009} were optimized (1) looking for higher stellar density towards the inner Galaxy, and (2) for detecting SiO masers which are hosted by O-rich AGB stars. It has been established that outside of the solar circle the AGB population contains a higher fraction of carbon rich stars~\citep{Lian2014a,Groenewegen2018,Lewis2020}, which will be verified using the SiO maser detections currently done by BAaDE in this region.

Finally, we have also noted two features in the source distribution presented in Figure~\ref{fig_Galactic}. First, the Outer Galactic region contains a lower number of confirmed cross-matches, that is expected from the BAaDE target selection which, in turn, comes from the MSX selection. Second, there is no notable difference in the spatial distribution of variable stars compared to the unclassified (non-variable) sources according to~{\gaia} DR2.

\begin{figure}
\begin{center}
\resizebox{\hsize}{!}{\includegraphics{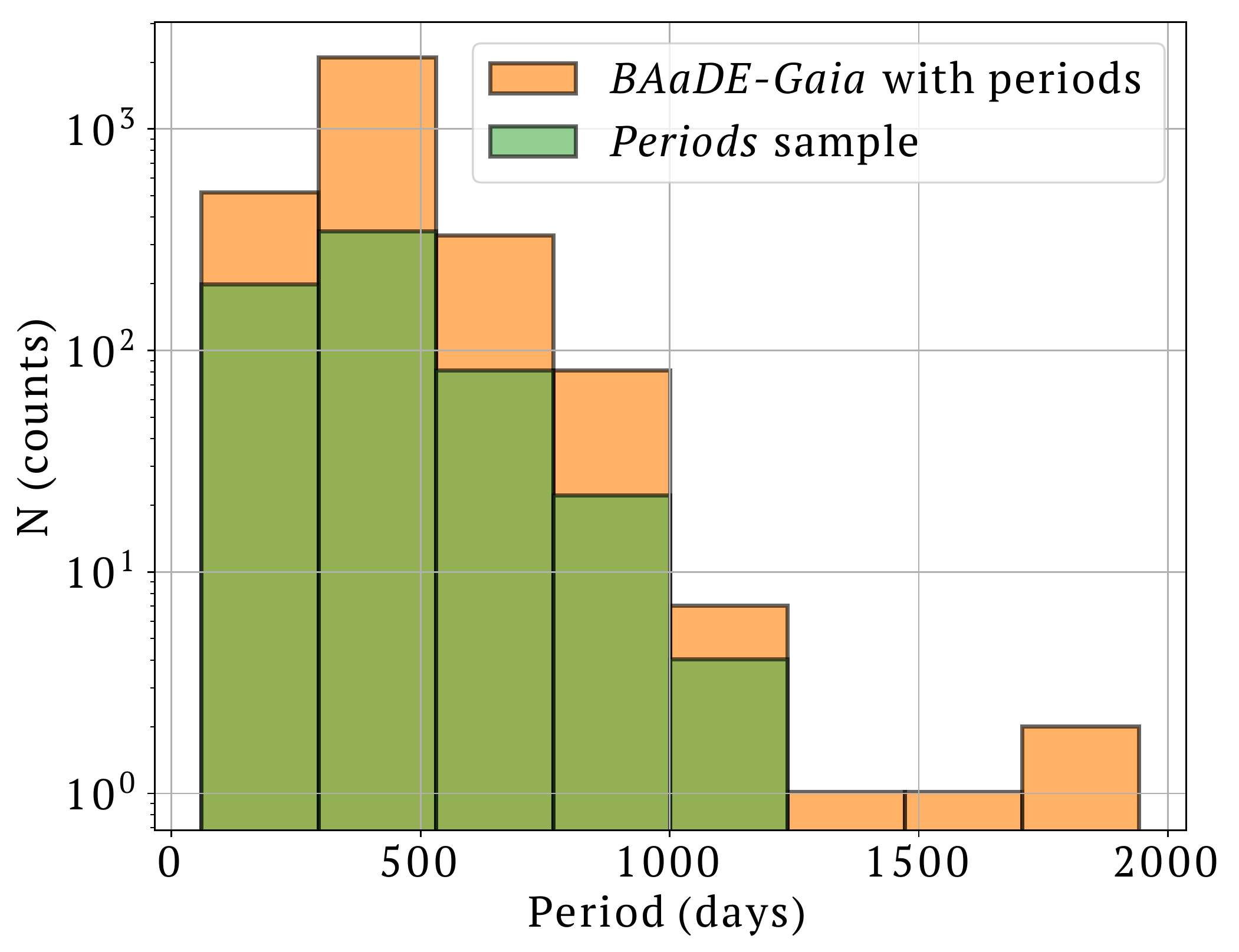}}
\caption{Period distributions obtained for variable stars in~{\gaia} DR2 for the {\it Gaia-BAaDE} and~{\local} samples (see Table~\ref{tab_filters} and Section~\ref{sec_periods}).}
\label{fig_histo_periods}
\end{center}
\end{figure}

\begin{figure}
\begin{center}
\resizebox{\hsize}{!}{\includegraphics{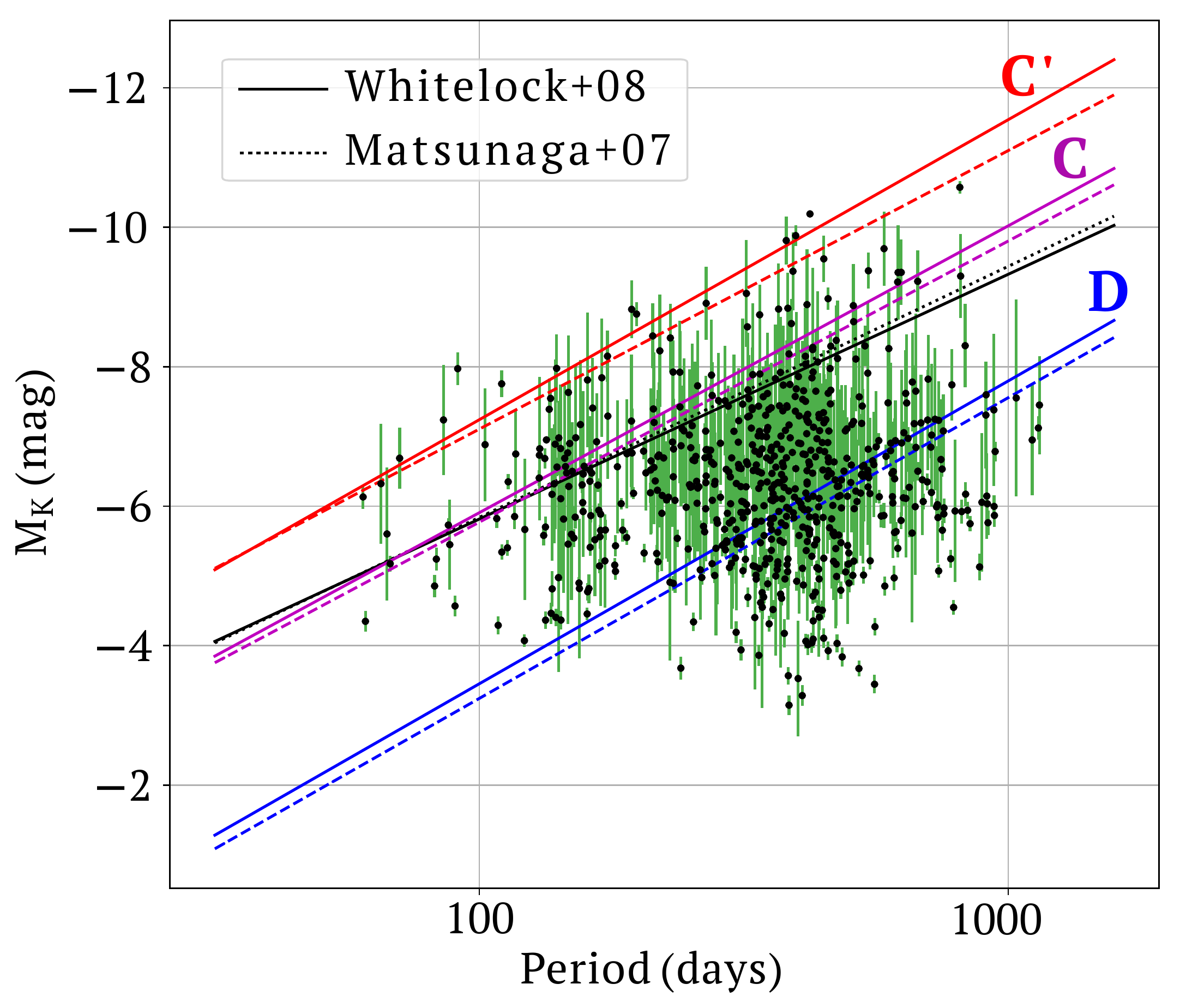}}
\caption{Period-Luminosity relations found for the variable stars within in the~{\local} sample. The sequences marked as $C$, $C$' and $D$ represent different known variability sequences, associated with distinct pulsation modes derived for the LMC based on~{\gaia} data~\citep{Lebzelter2019}. Note that the sequences were corrected for a distance modulus of 18.49 mag. Period-Luminosity relations reported for LMC using other surveys are also shown~\citep{Matsunaga2006,Whitelock2008}.}
\label{fig_period_lum}
\end{center}
\end{figure}

%\vspace{4mm}
\subsection{Period-Luminosity relations}
\label{sec_periods}

Accurately determined periods have been the means to distinguish variable stars but particularly pulsating variables (LPV, Cepheids, RR Lyrae, RV Tauri, etc.) within~{\gaia}~\citep{Eyer2019}. Figure~\ref{fig_histo_periods} displays the distribution of periods available from {\it Gaia} for sources in the {\it local} sample. It can be noted that the whole sample contains LPV stars with a wide range of periods, but that those within the solar neighborhood are restricted to stars with periods of $\lesssim$ 1250 days, presumably Mira or Semi-regular variables, as classified by~{\gaia}~\citep{Molnar2018,Mowlavi2018a}.

It has been established that Period-Luminosity (P-L) relations are a very powerful tool to distinguish AGB stars of different natures~\citep{Wood1999,Ita2004,Lebzelter2019}. By recognising that Mira variables pulsate dominantly in the fundamental mode, they can be promising candidates for distance determinations of remote galaxies, using empirical relations based on the LMC~\citep{Whitelock2008}.
With 2MASS $K$-magnitudes, {\it Gaia} DR2 parallaxes, extinction maps and periods for a sub-sample of the {\it local} sample ({\it period} sample), we are able to make a comparison of the BAaDE targets with previously studied variable stars.

In Figure~\ref{fig_period_lum}, we present the P-L relation for those BAaDE stars in the~{\local} sample with measured {\it Gaia} periods ({\it period} sample), where there is a spread in the magnitude, resulting from uncertainties in apparent $K$ magnitude, extinction and distance (indicated by the error bars) and IR variability (not indicated). A comparison is made of the P-L distribution with known variability sequences, associated with distinct pulsation modes, that have been derived from~{\gaia} DR2 data for LMC populations as discussed by \cite{Lebzelter2019}. These sequences have been transformed to $\rm{M_K}$, using the LMC distance modulus in that work (18.49 mag). Moreover, the established P-L relations for Miras from \cite{Whitelock2008} and~\cite{Matsunaga2006} in the LMC are added.

The {\local} sample appears to be much flatter than the empirical P-L relations for the LMC, and it is clear that most of the stars in the~{\local} sample fall below the LMC P-L relations~\citep{Whitelock2008,Matsunaga2006}. This is likely related to differences between the stellar samples used in Figure~\ref{fig_period_lum}. We preferably select the closer, less luminous AGB stars, when we make our~{\local} sample, while in the LMC the sample is (1) biased towards the most luminous stars and (2) have a different stellar metallicity. Therefore, it is possible that the LMC P-L relations previously related are missing a big, low-luminous clump that we are reporting. In fact, including more initial high-mass sources would add more sources to the upper right of Figure~\ref{fig_period_lum} \citep[see e.g.,][]{Vassiliadis1993}.

Using the analysis by~\cite{Lebzelter2019}, it is possible to further interpret Figure~\ref{fig_period_lum}. At short periods one can identify stars associated with sequence $C$, while at the most extreme long periods most star lie closer to sequence $D$. Supposedly both these sequences are being traced by low mass, oxygen rich Miras. At the intermediate periods, where there is the highest density of objects, there is no clear distinction between the two sequences. In \cite{Lebzelter2019}, the corresponding objects are mostly (extreme) carbon rich Miras. In this sense, we have already confirmed carbon stars in BAaDE sample based on IR color-cuts and detection rates~\citep[][]{Lewis2020b,Lewis2020}, but we plan on analyzing the observational results ---including the implications in the P-L diagram--- in the subsequent paper. Moreover, at longer periods and lying on sequence $D$, we typically find stars with a mass slightly over the solar mass and ages below 1 Gyr~\citep{Grady2019}. 

\section{Conclusions}
We have cross-matched the BAaDE target list, which consists of 28,062 IR sources mainly preselected from the MSX colors at latitudes $|b| < 5$\adeg{} to match evolved stars in the inner Galaxy~\citep{veen1988,Sjouwerman2009}, with the~{\gaia} DR2 catalogue~\citep{Brown2018}, finding 20,111 cross-matches. The cross-match was made using a conservative radius of 3\arcsec~around the MSX position which has a positional accuracy of 2\arcsec~\citep{Price1995}. One third of the BAaDE target list was not detected in~{\gaia} DR2, these sources correlate with lines of sight of high optical extinction in the Galactic plane. From the 20,111 cross-matched sources, stars with accurate parallax estimates and within a 2 kpc radius around the Sun (where we can obtain accurate extinction maps) were selected after removing YSOs and MS/RGB stars. The remaining 1,812 stars constitute our~{\local} sample, representing a foreground population of evolved stars in the Galactic plane. Among the~{\local} sample, the~{\gaia} DR2 shows large amplitude variability for 898 stars that have been classified as Mira variables~\citep{Mowlavi2018a}, of which another 649 have period estimates.

Using IR and optical data for this sample, we have characterized the evolved stellar population around the Sun in terms of spatial, variability, bolometric, and period-luminosity distributions. The population of evolved stars close to the Sun displays the following features:

\begin{enumerate}

\item The absolute magnitude distribution at $K$-band peaks at $-6.3\pm1.2$ mag with a spread of approximately 4 mag around the peak for the stars classified by~{\gaia} as variables. While the brightest sources are consistent with the expected luminosities for optically identified Mira variables, it is clear that our sample, at distances $<$2 kpc, mainly contains moderate luminosity variables. 

\item Using extinction and bolometric corrections from the literature, we are able to estimate bolometric magnitudes for the~{\local} foreground Galactic sample. The distribution peaks at $-$3.2 with a width of 1.2 mag ($1,500^{+3,000}_{-500} \ L_\odot$). This peak is at fainter magnitudes than that obtained for Miras in the LMC~\citep{Whitelock2008} and also at a lower value than inferred for the inner Galaxy~\citep{Trapp2018}. Although variability and uncertainties in the extinction and bolometric corrections are important, we argue that the main reason is the selection of faint, but nearby, sources that can be identified in the optical regime. 

\item By applying variability filters such as the amplitude index, it is possible to restrict the sample to LPV stars in the solar neighborhood. This filtering has shown to reproduce a narrower $K$-magnitude distribution similar to what we obtained after removing YSOs and MS/RGB stars.

\item Given that our samples are severely affected by interstellar extinction at~{\gaia} wavelengths, we have found that the distribution of moderately luminous evolved stars in the solar neighborhood disk ($r<2$ kpc) seem to be not correlated with respect to the location of major Galactic structures in the region. This confirms that the BAaDE target selection traces an old, dynamically relaxed stellar population.

\item For those BAaDE objects that have {\it Gaia} periods, we are able to associate these with fundamental mode and first overtone pulsation sequences. The BAaDE foreground population contains moderate mass Mira variables. Among the targets in the sample, carbon rich LPV stars also seem to be abundant. 

\end{enumerate}

Overall we conclude that the BAaDE targets are ---as discussed--- predominantly made up of LPVs, optically detectable Miras and carbon stars. The IR selection also picks up lower luminosity objects within 2 kpc from the Sun. The sample of evolved stars at these distances is mostly made up of AGB stars of moderate luminosity. To understand the nature of stars that make up the BAaDE sample in the inner Galaxy, advanced statistical methods that can use more uncertain~{\gaia} data combined with metallicity information would be required. In the subsequent paper, we will analyze the kinematics of the~{\local} sample by using a preliminary catalog of $\sim$17,000 sources, which corresponds for $\sim$70$\%$ of the VLA and $\sim$20$\%$ of the ALMA targets that have been already observed and analyzed as part of the BAaDE collaboration.

\acknowledgments
The authors sincerely thank the anonymous referee for making valuable suggestions that genuinely improved the paper. The BAaDE project is funded by National Science Foundation Grant 1517970/1518271. The National Radio Astronomy Observatory is a facility of the National Science Foundation operated under cooperative agreement by Associated Universities, Inc. M.C.S. is funded by the Heising-Simons Foundation under grant $\#$2018-0911. M.O.L received support for this work by the NSF through the Grote Reber Fellowship Program administered by Associated Universities, Inc./National Radio Astronomy Observatory. This research has made use of the NASA/IPAC Infrared Science Archive, which is operated by the Jet Propulsion Laboratory, California Institute of Technology, under contract with the National Aeronautics and Space Administration. This publication makes use of data products from the Two Micron All Sky Survey, which is a joint project of the University of Massachusetts and the Infrared Processing and Analysis Center/California Institute of Technology, funded by the National Aeronautics and Space Administration and the National Science Foundation. This work also has made use of data from the European Space Agency mission~{\gaia}, processed by the~{\gaia} Data Processing and Analysis Consortium (DPAC). Funding for the DPAC has been provided by national institutions, in particular the institutions participating in the~{\gaia} Multilateral Agreement. 

\bibliography{BAaDE-Gaia.bib}{}
\bibliographystyle{aasjournal}

\end{document}